\documentstyle[12pt]{article}

\newcommand{\sect}[1]{\setcounter{equation}{0}\section{#1}}

\newcommand{\EQ}{\begin{equation}}
\newcommand{\EN}{\end{equation}}
\newcommand{\bea}{\begin{eqnarray}}
\newcommand{\ena}{\end{eqnarray}}

\renewcommand{\a}{\alpha}
\renewcommand{\b}{\beta}

\renewcommand{\d}{\delta}

\newcommand{\pa}{\partial}

\newcommand{\G}{\Gamma}
\newcommand{\D}{\Delta}

\renewcommand{\l}{\lambda}
\renewcommand{\L}{\Lambda}

\newcommand{\r}{\rho}

\renewcommand{\S}{\Sigma}

\renewcommand{\o}{\omega}

\begin{document}
\def\bq{\begin{quote}}
\def\eq{\end{quote}}
\topmargin -1.2cm
\oddsidemargin 5mm

\renewcommand{\Im}{{\rm Im}\,}
\newcommand{\NP}[1]{Nucl.\ Phys.\ {\bf #1}}
\newcommand{\PL}[1]{Phys.\ Lett.\ {\bf #1}}
\newcommand{\NC}[1]{Nuovo Cimento {\bf #1}}
\newcommand{\CMP}[1]{Comm.\ Math.\ Phys.\ {\bf #1}}
\newcommand{\PR}[1]{Phys.\ Rev.\ {\bf #1}}
\newcommand{\PRL}[1]{Phys.\ Rev.\ Lett.\ {\bf #1}}
\newcommand{\MPL}[1]{Mod.\ Phys.\ Lett.\ {\bf #1}}
\renewcommand{\thefootnote}{\fnsymbol{footnote}}

\newpage
\begin{titlepage}
\begin{flushright}
IFUM 426/FT\\
hepth@xxx/9207018\\
\end{flushright}
\vspace{2cm}
\begin{center}
{\bf{{\large QUANTUM CONSERVED CURRENTS IN}}}\\
\vspace{.1in}
{\bf{{\large   SUPERSYMMETRIC TODA THEORIES}}} \\
\vspace{1.5cm}
{A. GUALZETTI},~~ { S. PENATI}~~ and~~ { D. ZANON} \\
\vspace{3mm}
{\em Dipartimento di Fisica dell'Universit\`{a} di Milano and} \\
{\em INFN, Sezione di Milano, Via Celoria 16, I-20133 Milano, Italy}\\
\vspace{1.1cm}
{{\bf{ABSTRACT}}}
\end{center}
\bq
We consider $N=1$ supersymmetric Toda theories which admit a fermionic
untwisted affine extension, i.e. the systems based on the $A(n,n)$, $D(n+1,n)$
and $B(n,n)$ superalgebras. We construct the superspace Miura trasformations
which allow to determine the W-supercurrents of the conformal theories and
we compute their renormalized expressions. The analysis of the renormalization
and conservation of higher-spin currents is then performed for the
corresponding
supersymmetric massive theories. We establish the quantum integrability of
these models and show that although their Lagrangian is not hermitian,
the masses of the fundamental particles are real,
a property which is maintained by one-loop
corrections. The spectrum is actually much richer, since the theories admit
solitons. The existence of quantum conserved higher-spin charges implies
that elastic, factorized S-matrices can be constructed.
\eq
\vfill
\begin{flushleft}
IFUM 426/FT\\
June 1992\\
\end{flushleft}
\end{titlepage}
\renewcommand{\thefootnote}{\arabic{footnote}}
\setcounter{footnote}{0}
\newpage

\sect{Introduction}
Whenever fermionic fields can be coupled consistently to their bosonic
partners, it is of interest to consider the supersymmetric version of the
theory. In particular this can be done for a class of
two-dimensional models known as Toda theories. The bosonic systems are
constructed from a Lie algebra and depending on whether
the algebra is affine or not, the resulting theories are massive or
conformally invariant, respectively. They possess an infinite number of
conservation laws which are not spoiled by
quantum anomalies \cite{b3,b4,b5}. This implies the existence of factorized,
elastic S-matrices \cite{b6}, which have been determined exactly for all
massive
theories based on simply-laced \cite{b7} as well as nonsimply-laced algebras
\cite{b8}.

Fermions can be added \cite{b1}
while maintaining the basic property of classical integrability: in this case
one has to construct a field theory starting from a Lie superalgebra \cite{b2}.
In general the introduction of fermions does not lead to
systems  invariant under supersymmetry transformations; indeed all
unitary fermionic Toda theories exhibit explicitly broken supersymmetry
\cite{b1}.
In order to obtain a supersymmetric model one has to consider those
superalgebras
which admit a purely fermionic simple root system. In this paper we focus
on these theories and among them we select the ones for which the untwisted
affine perturbation can be implemented also in a supersymmetric way.
Leaving aside the exceptional superalgebras $D(2,1;\a)$, we consider
the $A(n,n)$, $ D(n+1,n)$ and $B(n,n)$ supersymmetric Toda theories and address
the issue of their quantum integrability. We show that both in the conformal
and in the massive cases there exist quantum higher-spin conserved currents.
In the massive cases this ensures the commutativity of the corresponding
charges
with the S-matrix; thus the scattering is purely elastic and factorizes into
two-body processes. At the Lagrangian level these theories are nonunitary,
nonetheless the particle and soliton mass spectrum is real and the
explicit construction of exact S-matrices could be attempted.

In Section 2 we briefly review basic notions about superalgebras and set the
$N=1$ superspace notation. Then we  construct the classical Miura operators
for the above mentioned supersymmetric theories. The specific examples of the
$D(2,1)$, $B(1,1)$ and $A(1,1)$ conformal theories are presented in Section 3,
where we obtain the corresponding classical W-supercurrents and compute their
renormalized expressions. In Section 4 we consider the fermionic affine
extension of these models and we explicitly construct the first higher-spin
conserved currents to all orders of perturbation theory. Finally in Section 5
we compute the particle mass spectra up to one-loop for the
$A^{(1)}(n,n)$, $D^{(1)}(n+1,n)$ and $B^{(1)}(n,n)$ theories with $n=1,2$.
In accordance with previous results for unitary models,
we find that in the $A$ and $D$ series, being the
roots all of the same length (simply-laced), the classical
mass ratios are maintained
at the quantum level, while for the nonsimply-laced $B$-theories this is not
true. Section 6 contains some final remarks and our conclusions. In the
Appendix we have collected further details on the choice of the fermionic
roots and of the corresponding generators
and explicitly constructed the weights of the fundamental representation for
the various superalgebras.

\sect{N=1 supersymmetric Toda theories}
Two-dimensional
integrable theories with $N=1$ supersymmetry can be constructed as Toda models
based on a purely fermionic root system of a Lie superalgebra \cite{b1}.

A rank $r$ Lie superalgebra ${\cal G}={\cal G}_0 + {\cal G}_1$ is a
graded algebra whose
Lie bracket is defined as $[a,b] = ab - (-1)^{deg(a)deg(b)} ba$, where
$deg(a) = 0$ if $a \in {\cal G}_0$ and $deg(a) = 1$ if $a \in {\cal G}_1$.
The generators $\vec{h}$, $e_i^{+}$, $e_j^{-}$ satisfy the graded commutation
relations
\EQ
[\vec{h},\vec{h}] = 0 \qquad [\vec{h},e_j^{\pm}] = \pm \vec{\a}_j e_j^{\pm}
\qquad
[e_i^{+},e_j^{-}\} = \delta_{ij} \vec{h} \cdot \vec{\a}_i
\label{C1}
\EN
where $\vec{h}$ are the $r$ generators of the Cartan subalgebra and
$e_j^{\pm}$ are the basis vectors associated to the simple roots
$\vec{\a}_j$, with $j=1, \dots,r$.
Depending on whether $e_j^{\pm}$
is an element of ${\cal G}_0$ or of ${\cal G}_1$,
 $\vec{\a}_j$
is referred to as an {\it even} or {\it odd} root, respectively.

Lie superalgebras have been classified by Ka$\check{c}$ \cite{b2}
on the basis of the associated
Cartan matrix. A distinguishing feature
is given by the fact that the system of simple roots (i.e. the
Dynkin diagram) is not unique (up to a Weyl transformation). In general it is
possible to define several unequivalent
sets which have a different content of fermionic roots:
they are related by generalized Weyl transformations associated to
fermionic roots \cite{b9}.

A classical Lie superalgebra ${\cal G}$ can be suitably extended and a new
infinite dimensional one emerges, the {\it affine }
Lie superalgebra. In particular the
untwisted affine extension ${\cal G}^{(1)}$ is realized by adding to a system
of simple roots of ${\cal G}$ the {\it lowest} root defined as
\EQ
\vec{\a}_0 =-
\sum_{i=1}^r q_i \vec{\a}_i
\EN
 where $q_i$ are the Ka$\check{c}$ labels.
In the following we will focus on untwisted affine Lie
superalgebras which admit a purely fermionic extended set of simple roots.
They are the untwisted extensions of:

\noindent
1) the unitary series $A(n,n) \equiv sl(n+1,n+1;\cal{C})$,
$n \geq 1$.
The superalgebra $A(n,n)$ is not simple; it possesses a
one--dimensional graded invariant subalgebra generated by the basis element
${\bf I}_{2n+2}$. $A(n,n)$ is then defined as $sl(n+1,n+1;\cal{C})$
factored by this subspace.

\noindent
2) The orthosymplectic series $D(n+1,n) \equiv Osp(2n+2,2n;\cal{C})$,
$n \geq 1$ and
$B(n,n)=Osp(2n+1,2n;\cal{C})$, $n \geq 1$. These superalgebras are simple.

\noindent
3) Among the exceptional superalgebras only $D(2,1;\a)$, $\a \neq 0,-1$,
have a fermionic Dynkin
diagram with the lowest root also fermionic. They are deformations of the
$D(2,1)$ superalgebra.

\noindent
We restrict our analysis to the series listed in 1) and 2).

The classical field equations for a supersymmetric Toda theory based on a Lie
superalgebra can be viewed as the zero curvature condition on a Lax connection
in $N=1$ superspace $(z, \bar{z}, \theta, \bar{\theta})$, where $z$, $\bar{z}$
are Minkowski space light--cone coordinates
\bea
z \equiv x^+ = \frac{1}{\sqrt{2}} (x^0 + x^1) \qquad \qquad
\bar{z}\equiv x^-= \frac{1}{\sqrt{2}}(x^0 - x^1)  \nonumber \\
\partial \equiv \partial_z = \frac{1}{\sqrt{2}} (\partial_0 + \partial_1)
\qquad \qquad \bar{\partial} \equiv \partial_{\bar{z}} = \frac{1}{\sqrt{2}}
(\partial_0 - \partial_1) \qquad \qquad \Box = 2 \partial \bar{ \partial}
\ena
and $\theta$, $\bar{\theta}$ are the (1,1) spinor coordinates.
Introducing the corresponding covariant derivatives
\EQ
D = \partial_{\theta} + i \theta \partial  \qquad \qquad
\bar{D} = \partial_{\bar{\theta}} - i \bar{\theta} \bar{\partial}
\EN
satisfying the commutation relations $\{D, \bar{D}\} = 0$, $D^2 =
i \partial$, $\bar{D}^2 = -i \bar{\partial}$, and denoting by
$\vec{\Phi}$ the superfields
\EQ
\Phi^a = \phi^a + \frac{1}{\sqrt{2}} \theta \psi^a + \frac{1}{\sqrt{2}}
\bar{\theta} \bar{\psi}^a + \theta \bar{\theta} F^a  \qquad \quad a=1,
\dots ,r
\EN
the Lax connection is a Lie superalgebra--valued gauge superfield \cite{b1,b10}
\EQ
U =  D \vec{\Phi} \cdot \vec{h} - \lambda \sum_{j=0}^r e_j^{+}
\qquad \qquad \quad
\bar{U} = -\frac{1}{2\l} \sum_{j=0}^r q_j e^{\vec{\a}_j \cdot \vec{\Phi}}
e_j^{-}
\label{4}
\EN
which is flat, i.e. it satisfies
\EQ
\bar{D} U + D \bar{U} + \{U,\bar{U}\} = 0  \label{50}
\EN
In eq.(\ref{4}) we have introduced the spectral parameter $\l$ and chosen
the principal gradation for the generators.

It is straightforward to check using the commutation relations in eq.(\ref{C1})
that the flatness condition in eq.(\ref{50})
coincides with the equations of
motion derivable from the supersymmetric Toda action
\EQ
S = \frac{1}{\b^2} \int d^2 z d^2 \theta \left[ D\vec{\Phi} \cdot \bar{D}
\vec{\Phi} + \sum_{i=0}^{r} q_i e^{\vec{\a}_i \cdot \vec{\Phi}} \right]
\label{1}
\EN
where  $d^2 z = dz d \bar{z}$ , $d^2 \theta = \bar{D} D$
and $\b$ is the coupling constant. In order to obtain a realization
of the supersymmetric massive systems based on the affine superalgebras
$D^{(1)}(n+1,n)$, $B^{(1)}(n,n)$
and $A^{(1)}(n,n)$ one substitutes in eq.(\ref{1})
the explicit expressions of the fermionic roots
given in the Appendix. The conformally invariant Toda theories
are
obtained correspondingly
by dropping the exponential term containing the lowest root
$\vec{\a}_0$.

At the classical level supersymmetric Toda field theories are integrable.
The existence of an infinite
number of conservation laws is
a consequence of the identification of the Toda field equations
with eq.(\ref{50}).
The most practical way to construct explicitly the conserved supercurrents is
via the Miura operator which is obtained,
like in the bosonic case \cite{b11,b12,b4,b13},
by considering the Lax pair associated to eq.(\ref{50})
\EQ
(D + U) \chi=0 \qquad \qquad (\bar{D}+ \bar{U}) \chi=0
\label{2}
\EN
In order to describe the affine and the conformal cases on equal footing,
it is convenient to multiply the
generator $e_0^+$ associated to the lowest root by a c--number $c$,
so that setting $c=0$ ($\l =1$)
we restrict our results to the conformal theory,
whereas $c=1$ gives the corresponding formulae for the affine theory.
The Miura operators for the $D(n+1,n)$ and $B(n,n)$ conformal cases have also
been constructed in refs. \cite{added}.

We start by constructing the Miura operator for the $D(n+1,n)$ theory.
Using the explicit expression for the generators given in eqs.(A.12,A.13,A.15)
of the Appendix, the
first Lax equation in eq.(\ref{2}) can be written as
\EQ
(D + D\vec{\Phi} \cdot \vec{h}) \chi = \l \L \chi
\label{5}
\EN
where $(D + D\vec{\Phi} \cdot \vec{h})$ and $\L \equiv
(\sum_{j=1}^{2n+1} e_j^+ + c e_0^+)$ are the $(4n+2) \times (4n+2)$
matrices
\bea
(D + D\vec{\Phi} \cdot \vec{h}) &=&
{\rm diag} \{D + D\Phi_1, \cdots ,
D+D\Phi_{n+1}, D-D\Phi_1, \cdots , D-D\Phi_{n+1}, \nonumber \\
{}~~~~~~&~&~ D+iD\Phi_{n+2}, \cdots,
D+iD\Phi_{2n+1},D-iD\Phi_{n+2}, \cdots ,D-iD\Phi_{2n+1} \} \nonumber \\
{}~~~~~&~&~~~~
\ena
and
\EQ
\L =
\left(\begin{array}{ccccccccccc}
{}~&~ & ~ & ~ & ~ & ~ & | & {\bf I}_{n} & ~ & ~ & {\bf O}_{n} \\
{}~&~ & ~ & ~ & ~ & ~ & | & 0 & \cdots & 0 & 1 \\
{}~&~ & ~ & {\bf O}_{2n+2} & ~ & ~ & | & c & \cdots & 0 & 0 \\
{}~&~ & ~ & ~ & ~ & ~ & | & {\bf O}_{n} & ~ & ~ & {\bf I}_{n} \\
--&--&--&--&--&--&-|-&--&--&--&-- \\
0 & -{\bf I}_{n} & ~ & 0 & \cdots & 0 & | & ~ & ~ & ~ & ~   \\
\cdots & ~ & ~ & \cdots & \cdots & \cdots & | & ~ & ~ & ~ & ~ \\
0 & ~ & ~ & 0 & \cdots & -1 & | & ~ & {\bf O}_{2n} & ~ & ~ \\
c & 0 & \cdots & {\bf I}_{n} & ~ & 0 & | & ~ & ~ & ~ & ~  \\
0 & \cdots & \cdots & ~ & ~ & \cdots & | & ~ & ~ & ~ & ~ \\
\cdots & 0 & \cdots & ~ & ~ & 0 & | & ~ & ~ & ~ & ~
\end{array} \right)
\EN
The vector $\chi$ in eq.(\ref{5}) has $(4n+2)$ superfield entries, the first
$2n+2$ obeying opposite statistics with respect to the others.
Thus eq.(\ref{5}) gives rise to the following set of
coupled equations
\bea
(D+D\Phi_j) \chi_j &=& \l \chi_{2n+2+j}\qquad
\qquad \qquad \qquad j = 1, \cdots ,
n \nonumber \\
(D+D\Phi_{n+1}) \chi_{n+1} &=& \l \chi_{4n+2} \nonumber \\
(D-D\Phi_1) \chi_{n+2} &=& c~\l \chi_{2n+3} \nonumber \\
(D-D\Phi_j) \chi_{n+1+j} &=& \l \chi_{3n+1+j}
{}~~~\qquad \qquad \qquad \quad j=2, \cdots
,n+1 \nonumber \\
(D+iD\Phi_{n+1+j}) \chi_{2n+2+j} &=& -\l \chi_{j+1} \quad ~\qquad
\qquad \qquad \quad ~
j=1, \cdots ,n-1 \nonumber \\
(D+iD\Phi_{2n+1}) \chi_{3n+2} &=& -\l (\chi_{n+1} + \chi_{2n+2}) \nonumber \\
(D-iD\Phi_{n+2}) \chi_{3n+3} &=& \l \chi_{n+2} + c~\l \chi_1 \nonumber \\
(D-iD\Phi_{n+1+j}) \chi_{3n+2+j} &=& \l \chi_{n+1+j} \quad
\qquad \qquad \qquad ~\quad
j=2, \cdots ,n
\ena
It is easy to see that one can eliminate
iteratively the $\chi_2, \cdots , \chi_{4n+2}$ superfields
and derive a single equation for $\chi_1$
\EQ
\D(\vec{\Phi}) \chi_1 = 4c~(-1)^{n} \l^{4n} D \chi_1
\EN
where
\bea
\D(\vec{\Phi}) &=&
(D-D\Phi_1)(D-iD\Phi_{n+2})(D-D\Phi_2)(D-iD\Phi_{n+3}) \cdots
(D-iD\Phi_{2n+1}) \nonumber \\
&~~&(D-D\Phi_{n+1}) \frac{1}{D} (D+D\Phi_{n+1})(D+iD\Phi_{2n+1})
\cdots (D+iD\Phi_{n+2})(D+D\Phi_1) \nonumber\\
{}~~~&~&~~~\label{miura1}
\ena
is the Miura operator for the $D(n+1,n)$ Toda theory.
We note that $\D(\vec{\Phi})$ can be written equivalently
in terms of
the weights of the fundamental representation of the algebra (see eq.(A.17) in
the Appendix)
\bea
\D(\vec{\Phi}) &=&
(D+\vec{\l}_{4n+2} \cdot D\vec{\Phi})(D+\vec{\l}_{4n+1}
\cdot D\vec{\Phi}) \cdots
\cdots (D+\vec{\l}_{2n+2} \cdot D\vec{\Phi})  \nonumber \\
&~~&\frac{1}{D} (D+\vec{\l}_{2n+1} \cdot D\vec{\Phi})(D+\vec{\l}_{2n}
\cdot D\vec{\Phi})
\cdots (D+\vec{\l}_{1} \cdot D\vec{\Phi})
\ena
In this case, in complete analogy with the corresponding Miura operator
for the bosonic Toda theory based on the $d_n$ Lie algebra \cite{b12,b13},
$\D(\vec{\Phi})$ is a pseudo differential operator.

We specialize now the first Lax equation in (\ref{2}) to the $B(n,n)$ system.
Using the results listed in eqs.(A.19,A.20,A.22) of the Appendix, we have
\bea
(D + D\vec{\Phi} \cdot \vec{h}) &=&
{\rm diag} \{D + D\Phi_1, \cdots ,
D+D\Phi_n, D-D\Phi_1, \cdots , D-D\Phi_n, D, \nonumber \\
{}~~~~~~~~~&~&~ D+iD\Phi_{n+1}, \cdots ,
D+iD\Phi_{2n},D-iD\Phi_{n+1}, \cdots ,D-iD\Phi_{2n} \} \nonumber\\
{}~~~&~&~~~
\ena
and $\L = \sum_{j=1}^{2n}e_j^+ + c e_0^+$, that is
\EQ
\L =
\left(\begin{array}{ccccccccccc}
{}~ & ~ & ~ & ~ & ~ & ~ & | & {\bf I}_n & ~ & {\bf O}_n & ~ \\
{}~ & ~ & {\bf O}_{2n+1} & ~ & ~ & ~ & | & c & 0 & \cdots & 0  \\
{}~ & ~ & ~ & ~ & ~ & ~ & | & {\bf O}_n & ~ & {\bf I}_n & ~ \\
--&--&--&--&--&--&-|-&--&--&--&-- \\
0 & -{\bf I}_{n-1} & ~ & 0 & \cdots & 0 & | & ~ & ~ & ~ & ~   \\
\cdots & ~ & ~ & \cdots & \cdots & \cdots & | & ~ & ~ & ~ & ~ \\
0 & 0 & \cdots & 0 & \cdots & -1 & | & ~ & {\bf O}_{2n} & ~ & ~ \\
c & 0 & \cdots & {\bf I}_n & ~ & 0 &| & ~ & ~ & ~ & ~  \\
0 & \cdots & \cdots & ~ & ~ & \cdots & | & ~ & ~ & ~ & ~ \\
\cdots & \cdots & \cdots & ~ & ~ & 0 & | & ~ & ~ & ~ & ~
\end{array} \right)
\EN
In this case the vector $\chi$ has the first $(2n+1)$ components with opposite
statistics with respect to the last $2n$.
The Lax equation is then equivalent to the system
\bea
(D+D\Phi_j) \chi_j &=& \l \chi_{2n+1+j} ~~\qquad \qquad \quad j = 1, \cdots ,
n \nonumber \\
(D-D\Phi_1) \chi_{n+1} &=& c~\l \chi_{2n+2} \nonumber \\
(D-D\Phi_j) \chi_{n+j} &=& \l \chi_{3n+j} \qquad ~~~~~ \qquad \quad j
=2, \cdots ,n \nonumber \\
D \chi_{2n+1} &=& \l \chi_{4n+1} \nonumber \\
(D+iD\Phi_{n+j}) \chi_{2n+1+j} &=& -\l \chi_{j+1} ~\quad \qquad \qquad \quad
j=1, \cdots ,n-1 \nonumber \\
(D+iD\Phi_{2n}) \chi_{3n+1} &=& -\l \chi_{2n+1} \nonumber \\
(D-iD\Phi_{n+1}) \chi_{3n+2} &=& \l \chi_{n+1} + c~\l \chi_1 \nonumber \\
(D-iD\Phi_{n+j}) \chi_{3n+1+j} &=& \l \chi_{n+j} \qquad \qquad \qquad \quad
j=2, \cdots ,n
\ena
from which $\chi_2, \cdots ,\chi_{4n+1}$ can be eliminated and we obtain
\EQ
\D(\vec{\Phi}) \chi_1 = 2c~(-1)^{n} \l^{4n} D\chi_1
\EN
where
\bea
\D(\vec{\Phi}) &=&
(D-D\Phi_1)(D-iD\Phi_{n+1})(D-D\Phi_2)(D-iD\Phi_{n+2}) \cdots
(D-D\Phi_n) \nonumber \\
&~~& (D-iD\Phi_{2n}) D(D+iD\Phi_{2n})(D+D\Phi_n)
\cdots (D+iD\Phi_{n+1})(D+D\Phi_1) \nonumber \\
&~~&~~  \label{miura2}
\ena
The operator in eq.(\ref{miura2})
generates the $W$--currents for the conformal $B(n,n)$ Toda field theory.
It can be written as
\EQ
\D(\vec{\Phi}) =
(D+\vec{\l}_{4n+1} \cdot D\vec{\Phi})(D+\vec{\l}_{4n} \cdot D \vec{\Phi})
\cdots \cdots
(D+ \vec{\l}_{2} \cdot D\vec{\Phi})(D+\vec{\l}_{1} \cdot D\vec{\Phi})
\EN
where $\vec{\l}_1, \cdots , \vec{\l}_{4n+1}$ are the weights of the
fundamental representation of the algebra (see eq.(A.23)).

Finally we consider the Miura transformation for the $A(n,n)$ case. The
generators of the algebra obtained by factoring out the invariant
subalgebra $\{{\bf I}_{2n+2}\}$ are given in eqs.(A.24,A.25,A.26)
of the Appendix.
We have
\bea
(D + D\vec{\Phi} \cdot \vec{h}) &=&
{\rm diag} \{D + D\Phi_1 +\cdots +D\Phi_n +iD\Phi_{n+1} +\cdots +
iD\Phi_{2n}, \nonumber \\
{}~~~~~~&~&~ D -D\Phi_1 + D\Phi_2+ \cdots +D\Phi_n+iD\Phi_{n+1}+\cdots
+iD\Phi_{2n}, \nonumber \\
{}~~~~~~&~&~ D - D\Phi_2 + D\Phi_3 +\cdots +D\Phi_n +iD\Phi_{n+2} +\cdots +
iD\Phi_{2n}, \nonumber \\
{}~~~~~&~&~ \cdots \cdots , D - D\Phi_n +iD\Phi_{2n},  \nonumber \\
{}~~~~~~&~&~ D + D\Phi_2 +\cdots + D\Phi_n +2iD\Phi_{n+1} +iD\Phi_{n+2}
+\cdots +iD\Phi_{2n}, \nonumber \\
{}~~~~~~&~&~ D+D\Phi_3 +\cdots + D\Phi_n +2iD\Phi_{n+2} +iD\Phi_{n+3}
+\cdots +iD\Phi_{2n}, \nonumber \\
{}~~~~~~&~& \cdots \cdots , D+2iD\Phi_{2n}, D \}
\ena
and
\EQ
\L = \sqrt{2}~
\left(\begin{array}{ccccc}
{}~ & ~ & | & ~ & ~ \\
{}~ & {\bf O}_{n+1} & | & {\bf I}_{n+1} & ~  \\
--&--&-|-&--&-- \\
0 ~ & {\bf I}_n & | & {\bf O}_{n+1} & ~ \\
c & 0 & | & ~ & ~
\end{array} \right)
\EN
The $(2n+2)$--dimensional vector $\chi$ contains $(n+1)$ fermionic and
$(n+1)$ bosonic superfields. The Lax equation
$(D + D\vec{ \Phi} \cdot \vec{h}) \chi = \l \L \chi$
gives rise to the system
\bea
(D+\sum_{k=1}^n D\Phi_k+i\sum_{k=n+1}^{2n} D\Phi_k) \chi_1 &=&
\sqrt{2} \l \chi_{n+2} \nonumber \\
(D - D\Phi_j +\sum_{k=j+1}^n D\Phi_k + i\sum_{k=n+j}^{2n} D\Phi_k )
\chi_{j+1} &=& \sqrt{2} \l \chi_{n+2+j}
\qquad \quad j = 1, \cdots , n \nonumber \\
(D + \sum_{k=j}^n D\Phi_k + 2iD\Phi_{n+j-1} +i\sum_{k=n+j}^{2n} D\Phi_k)
\chi_{n+j} &=& \sqrt{2} \l \chi_j
\qquad \quad \qquad j=2, \cdots ,n+1 \nonumber \\
D\chi_{2n+2} &=& c~\sqrt{2}  \l \chi_1
\label{7}
\ena
Again it can be reduced to
\EQ
\D(\vec{\Phi}) \chi_1 =c (\sqrt{2} \l)^{2n+2} \chi_1
\EN
where
\bea
\D(\vec{\Phi}) &=&
D(D- D\Phi_n+iD\Phi_{2n})(D + 2iD\Phi_{2n}) (D- D\Phi_{n-1} +D\Phi_n
+iD\Phi_{2n-1} + iD\Phi_{2n}) \nonumber \\
&~& \cdots \cdots
(D+ D\Phi_2+\cdots + D\Phi_n +2iD\Phi_{n+1} +iD\Phi_{n+2} +\cdots +
iD\Phi_{2n} ) \nonumber \\
&~& (D+ D\Phi_1+\cdots + D\Phi_n +iD\Phi_{n+1} + \cdots + iD\Phi_{2n})
\label{miura3}
\ena
gives the Miura operator for
the $A(n,n)$ Toda field theory. This operator has the general form
\EQ
\D(\vec{\Phi}) =
(D+\vec{\l}_{2n+2} \cdot D\vec{\Phi})(D + \vec{\l}_{2n+1} \cdot D\vec{\Phi})
\cdots \cdots (D+\vec{\l}_1 \cdot D \vec{\Phi})
\EN
in terms of the weights of the fundamental representation of the superalgebra
(see eq.(A.29)). We note that the results in eqs.(2.22,2.28) are in agreement
with the statement in Ref.\cite{b10}.

{}From the knowledge of the differential operators in
eqs.(2.15,2.21,2.27) one easily obtains the currents
of the conformal Toda theories. We construct them explicitly in the next
section.

\sect{$W$--supercurrents of the conformal theories}

As we have mentioned, the general action which describes the
supersymmetric conformally invariant Toda field theories
can be obtained from eq.(\ref{1})
by neglecting the interaction term associated to
the lowest root $\vec{\a}_0$.
For such systems the existence of classically
conserved higher--spin supercurrents follows from the fact that
the Toda field equations are the integrability conditions
of the linear problem
in eq.(\ref{2}). Indeed, in the previous section the first of these
two equations has been used to construct the Miura operator
$\D(\vec{\Phi})$ which, for the $B(n,n)$ and $A(n,n)$ theories,
can be written as
\EQ
\D(\vec{\Phi}) = D^{d} + \sum_{i=0}^{d-1} W^{(\frac{d-1}{2}-\frac{i}{2})} D^i
\label{13}
\EN
where $d$ is the dimension of the fundamental representation of the
superalgebra.
For the $D(n+1,n)$ theory the Miura operator in eq.(\ref{miura1})
is non local and an identity like (\ref{13}) holds for the local part.
In addition, because of the $\frac{1}{D}$
factor, $d$ equals the dimension of the
fundamental representation decreased by one.

For the $D(n+1,n)$ and $B(n,n)$ theories, using the Toda field equations,
one can show that
\EQ
[\bar{D}, \D(\vec{\Phi})] = 0
\label{holom}
\EN
Therefore the $W^{(s)}$ coefficients in eq.(\ref{13})
provide a set of classical
superholomorphic spin--$s$ currents
\EQ
\bar{D} W^{(s)} = 0
\label{holom1}
\EN
which ensure the classical integrability of these Toda systems.
For the $A(n,n)$ case the equation (\ref{holom}) is not valid, however
the $W^{(1)}$ current satisfies (\ref{holom1}). Since the $W^{(1)}$ current
is proportional to the stress--energy tensor all the Toda theories under
consideration describe $N=1$ superconformal models.

The renormalization of the conservation laws in (\ref{holom1}) can be studied
following the procedure described in Ref.\cite{b14}.
The calculation is performed using massless perturbation theory, which is
best suited for an all--loop analysis \cite{b5}. We define the
quantum Lagrangian by normal ordering the exponentials (no
ultraviolet divergences), which are then treated as interaction terms.
Potential anomalies in the quantum conservation laws would
arise from {\em local} terms in
\EQ
\bar{D}_Z \left \langle W^{(s)}(Z,\bar{Z}) \right \rangle \equiv \bar{D}_Z
\left \langle W^{(s)}(Z,\bar{Z})
\exp \left(\frac{i}{\b^2}
\int d^2w d^2 \theta' {\cal L}_{int} \right) \right \rangle _0
\label{9}
\EN
where $(Z,\bar{Z}) \equiv (z,\theta,\bar{z},\bar{\theta})$.
We introduce massless superspace propagators
\EQ
\left \langle
\Phi_i(Z,\bar{Z}) \Phi_j(0,0)\right \rangle =
- \d_{ij} \frac {\b^2}{4\pi} \bar{D} D[log(2z \bar{z})
\d^{(2)}(\theta)]
\EN
and compute Wick contractions of the currents with the exponential in
eq.(\ref{9}).
A spin--$s$ current consists in a sum of terms with $2s+1$ spinor derivatives
acting on the fields. Standard superspace techniques allow to perform
first the integration on the $\theta$--variable: in a given loop one
reduces the number of covariant spinor derivatives from the various terms in
the current and from the superspace propagators, using their commutation
relations, to at most one $D$ and one $\bar{D}$ times a number of
space--time derivatives. One obtains then a nonvanishing contribution if
one ends up with exactly one $D$ and one $\bar{D}$, in which case
\EQ
\d^{(2)}(\theta -\theta') \bar{D} D\d^{(2)}(\theta -\theta') =
\d^{(2)}(\theta -\theta')
\EN
Moreover since we are interested in determining only local contributions,
it is sufficient to
expand the exponential in eq.(\ref{9}) to first order in ${\cal L}_{int}$.
In fact, once the $D$--algebra has been performed we obtain terms of the form
\EQ
\bar{D}_Z \int d^2 w {\cal A}(z,\bar{z},\theta, \bar{\theta})
\bar{D}_Z \frac {1} {(z-w)^n}  {\cal{B}}(w, \bar{w}, \theta, \bar{\theta})
\EN
where ${\cal A}$, ${\cal B}$ are products of superfields and their
$D$--derivatives.
Now, using the prescription
\EQ
\bar{D}_Z \bar{D}_Z \frac{1}{(z-w)^n}= -i \bar{\pa}_z \frac{1}{(z-w)^n}
=\frac{2\pi}{(n-1)!} \pa^{n-1}_w \d^{(2)}(z-w) \label{int}
\EN
we produce local contributions
that we will try to cancel by renormalizing the classical $W^{(s)}$-currents,
i.e. modifying them by coupling constant dependent terms.
Thus for example, if we consider a typical contribution from a spin--$1$
current, like the energy--momentum tensor, we obtain
\newpage
\bea
&& \bar{D}_Z \left \langle D \Phi \pa \Phi
\left( \frac{i}{\b^2} \int d^2w d^2 \theta' e^{\Phi} \right)
\right \rangle    \nonumber \\
{}~~~~&& \leadsto i \bar{D}_Z \int d^2w
\left[ D \Phi \bar{D} \frac {-1}{4\pi (z-w)} D e^{\Phi}+
\pa \Phi \bar{D} \frac{i}{4\pi (z-w)} e^{\Phi}
- \frac{i \b^2}{16\pi^2} \bar{D} \frac{1}{(z-w)^2} e^{\Phi} \right]
\nonumber \\
{}~~~~&&\leadsto \frac{i}{2} D \Phi De^{\Phi} -\frac{1}{2} \pa \Phi e^{\Phi} -
\frac{\b^2}{8\pi} \pa e^{\Phi}
\ena
where in the last equality we have kept the first term, which actually
vanishes, for pedagogical reasons.

In the next subsections we construct the $W$--currents for the
simplest examples of superconformal Toda theories, namely the $D(2,1)$,
$B(1,1)$ and $A(1,1)$ theories. We compute first their
classical expressions as derived from the Miura
operator and  obtain then their exact quantum corrected version.
\vskip 10pt
\noindent
{\bf 1) ${\bf D(2,1)}$ Toda theory}

The $D(2,1)$ superalgebra has rank $3$ and
we can express the fermionic simple roots as $3$--dimensional
vectors (cfr. eq.(A.12))
\bea
\vec{\a}_1&=& \vec{\varepsilon}_1 - \vec{\delta}_1 =(1,0,-i) \nonumber \\
\vec{\a}_2&=&\vec{\delta}_1 - \vec{\varepsilon}_2 =(0,-1,i) \nonumber \\
\vec{\a}_3&=&\vec{\delta}_1 + \vec{\varepsilon}_2=(0,1,i)
\label{root1}
\ena
The Toda action is written in terms of $3$ scalar superfields
\EQ
S= \frac{1}{\b^2} \int d^2 z d^2 \theta \left[ D \vec{\Phi} \cdot \bar{D}
\vec{\Phi} + e^{\Phi_1 -i\Phi_3} + e^{-\Phi_2 +i\Phi_3} + e^{\Phi_2
+i\Phi_3} \right]
\label{14}
\EN
It is symmetric under $\Phi_2 \rightarrow -\Phi_2$;
this will give rise to a corresponding definite parity
of the currents.

Specializing the Miura operator in eq.(\ref{miura1}) to the $n=1$ case we
obtain
\bea
\D(\vec{\Phi}) &=& (D-D\Phi_1)(D-iD\Phi_3)(D-D\Phi_2)\frac{1}{D}
(D+D\Phi_2)(D+iD\Phi_3)(D+D\Phi_1) \nonumber \\
&=& D^5 + \sum_{i=0}^2 W^{(2-\frac{i}{2})}~D^i + F^{(1)} \frac{1}{D} F^{(1)}
\label{10}
\ena
where we have defined
\EQ
F^{(1)} \equiv (D-D\Phi_1)(D-iD\Phi_3)(D-D\Phi_2)~{\bf 1}
\EN
The local part of eq.(\ref{10}) gives the following superholomorphic currents
\bea
W^{(1)} &=& -iD \Phi_1 \partial \Phi_1 - iD\Phi_2 \partial \Phi_2 -
iD\Phi_3 \partial \Phi_3
+ 2iD \partial \Phi_1 + D \partial \Phi_3  \nonumber \\
W^{(\frac{3}{2})} &=& -i {\partial}^2 \Phi_3 +iD\Phi_1 D \partial \Phi_1 -
iD\Phi_2 D \partial
\Phi_2 - (\partial \Phi_3)^2 - i D\Phi_3 D \partial \Phi_3 \nonumber \\
&~& +2 D \Phi_1 D \partial \Phi_3 - 2i D\Phi_1 D\Phi_2 \partial \Phi_2 -
2i D\Phi_1
D\Phi_3 \partial \Phi_3 + 2 D\Phi_2 \partial \Phi_2 D \Phi_3  \nonumber \\
W^{(2)} &=& \frac12 D(W^{(\frac{3}{2})} + D W^{(1)})
\label{100}
\ena
whereas from the nonlocal term we obtain an additional spin--$1$ holomorphic
current
\EQ
F^{(1)} = -i D\pa \Phi_2 +iD\Phi_1 \pa \Phi_2 - iD(D\Phi_2 D\Phi_3) +i D\Phi_1
D\Phi_2 D\Phi_3
\label{101}
\EN
In particular $W^{(1)}$ can be written as
\EQ
W^{(1)} = -iD \vec{\Phi} \cdot \partial \vec{\Phi} + 2i \vec{\rho}
\cdot D \partial \vec{\Phi}
\label{12}
\EN
where $\vec{\rho}$ is the Weyl vector for the $D(2,1)$ superalgebra (see
eq.(A.16) in the Appendix).
It coincides with the improved stress--energy tensor of the
theory.

We turn now to the study of the quantum extension of the classical conserved
currents $W^{(1)}$, $F^{(1)}$ and $W^{(\frac{3}{2})}$.
We do not need to compute
separately the renormalization of $W^{(2)}$ since it depends linearly on
$W^{(1)}$ and $W^{(\frac{3}{2})}$.

Inserting $W^{(1)}$ in eq.(\ref{9}) and searching for potential
anomalous terms, it is easy to check that in this case no
anomalies appear, so that the classical expression in eq.(\ref{12}) is
not renormalized.
This result agrees with the general statement \cite{b10} based on the study of
the conformal dimensions of the interaction exponentials, according to
which the quantum corrections to the stress--energy tensor
can be obtained by a renormalization of the Weyl vector
\EQ
\vec{\rho} = \frac12 \sum_i \vec{\nu}_i ~~~~\longrightarrow ~~~~
\vec{\rho}_q = \sum_i \frac{1}{2}(1 + \frac{\b^2}{4\pi}
\vec{\a}_i^2)\vec{ \nu}_i
\label{16}
\EN
where the vectors $\vec{\nu}_i$ are dual to the fermionic
simple roots.
Since in our case all the roots in eq.(\ref{root1}) have vanishing square, no
renormalization is required.
A simple calculation shows that also the $F^{(1)}$ current is not renormalized.

The corresponding analysis for the spin--$\frac{3}{2}$ current is
somewhat more complicated. The most efficient way to compute
the quantum corrections is to consider a
linear combination of terms as in eq.(\ref{100}) with arbitrary coefficients
\bea
W^{(\frac{3}{2})} &=& a \pa^2 \Phi_3 + b D\Phi_1 D \pa \Phi_1
+c D\Phi_2 D\pa \Phi_2 + d (\pa \Phi_3)^2
 \nonumber \\
&~~&+ e D\Phi_3 D\pa \Phi_3 + f D\Phi_1 D \pa \Phi_3 +g D \Phi_1 D \Phi_2
\pa \Phi_2
\nonumber \\
&~~&+ h D \Phi_1 D \Phi_3 \pa \Phi_3 + j D \Phi_2 \pa \Phi_2 D \Phi_3
\label{W32}
\ena
The unknowns $a,b,\cdots,j$ are then determined by requiring
absence of anomalies in the quantum conservation law. To this end we
proceed as described at the beginning of this section and compute all
the {\em local} contributions which arise from Wick contracting the expression
in eq.(\ref{W32}) with
the interaction Lagrangian. We obtain
($\a \equiv \frac{\b^2}{2 \pi}$)
\bea
&& \bar{D} \left \langle W^{(\frac{3}{2})}
\left( \frac{i}{\b^2} \int d^2w d^2 \theta' e^{\Phi_1-i\Phi_3} \right)
\right \rangle \nonumber \\
&& \leadsto \left(
\left[-\frac{1}{2} a +\frac{1}{2}b +\frac{i\a}{4}d -\frac{i}{2}f -
\frac{\a}{4}h
 \right] D\Phi_1 \pa \Phi_1 \right. \nonumber \\
&& \left.
+\left[ \frac{i}{2}a-\frac{i}{2}b+(-1+\frac{\a}{4})d -\frac{1}{2}f
+\frac{i}{2}(-1+\frac{\a}{2})h
\right] D\Phi_1 \pa \Phi_3
\right. \nonumber \\
&& \left.
+\left[-\frac{1}{2}a -\frac{1}{2}b +\frac{i\a}{4}d  \right] D\pa \Phi_1
+\left[\frac{i}{2}a +\frac{\a}{4}d +\frac{i}{2}e-\frac{1}{2}f
 \right] D \pa \Phi_3
\right. \nonumber \\
&&\left.
+\left[ -\frac{1}{2}g-\frac{i}{2}j \right] D \Phi_2 \pa \Phi_2
+\left[\frac{1}{2}a +i(1-\frac{\a}{4})d-\frac{1}{2}e+
\frac{1}{2}(-1+\frac{\a}{2})h
 \right] D\Phi_3 \pa \Phi_3 \right. \nonumber \\
&& \left.
+\left[\frac{i}{2}a+\frac{\a}{4}d-\frac{i}{2}e+\frac{i\a}{4}h
\right] D\Phi_3 \pa \Phi_1    \right) e^{\Phi_1-i\Phi_3}
\ena
and
\bea
&& \bar{D} \left \langle W^{(\frac{3}{2})}
\left( \frac{i}{\b^2} \int d^2w d^2 \theta' e^{\Phi_2+i\Phi_3} \right)
\right \rangle \nonumber \\
&& \leadsto \left(
\left[ \frac{i}{2}a +\frac{i}{2}c +(1-\frac{\a}{4})d -\frac{\a}{4}j
 \right] D\Phi_2 \pa \Phi_3
+\left[\frac{1}{2}a-\frac{1}{2}c+\frac{i\a}{4}d \right] D \pa \Phi_2 \right.
\nonumber \\
&& \left.
+\left[\frac{i}{2}f+\frac{1}{2}(1+\frac{\a}{2})g -\frac{\a}{4}h
\right] D \Phi_1 \pa \Phi_2
+\left[-\frac{1}{2}f+\frac{i\a}{4}g+\frac{i}{2}(1-\frac{\a}{2})h
\right] D\Phi_1 \pa \Phi_3 \right. \nonumber \\
&& \left.
+\left[ \frac{i}{2}a-\frac{\a}{4}d+\frac{i}{2}e-\frac{1}{2}(1+\frac{\a}{2})j
 \right] D \Phi_3 \pa \Phi_2 \right.
\nonumber \\
&&\left.
+\left[\frac{1}{2}a+\frac{1}{2}c+\frac{i\a}{4}d+\frac{i}{2}(1+\frac{\a}{2})j
\right] D \Phi_2 \pa \Phi_2 \right. \nonumber \\
&& \left.
+\left[-\frac{1}{2}a+i(1-\frac{\a}{4})d-\frac{1}{2}e-\frac{i\a}{4}j
\right] D\Phi_3 \pa \Phi_3
+\left[\frac{1}{2}g-\frac{1}{2}h \right] D\Phi_1 D\Phi_2 D\Phi_3 \right.
\nonumber \\
&&\left.
+\left[\frac{i}{2}a-\frac{\a}{4}d-\frac{i}{2}e \right]
D \pa \Phi_3    \right) e^{\Phi_2+i\Phi_3}
\ena
The third exponential needs not be considered because of the
$\Phi_2$--symmetry of the Lagrangian.
The $W^{(\frac{3}{2})}$ current will satisfy the conservation equation
(\ref{holom1}) if the coefficients of the various terms separately vanish.
This leads to a set of equations for
$a,b,\dots ,j$ which can be solved {\em nontrivially}. Thus we
obtain, up to an overall normalization factor, the {\em quantum}
spin-$\frac{3}{2}$ conserved current
\bea
W^{(\frac{3}{2})} &=& -i {\partial}^2 \Phi_3 +i \left(1-\frac{\b^2}{4\pi}
\right) D\Phi_1 D \partial \Phi_1 - i \left(1+\frac{\b^2}{4\pi} \right)
D\Phi_2 D \partial \Phi_2 \nonumber \\
&~~& - (\partial \Phi_3)^2 - i\left(1+\frac{\b^2}{4\pi}\right)
D\Phi_3 D \partial \Phi_3
+ 2 D \Phi_1 D \partial \Phi_3 - 2i D\Phi_1 D\Phi_2 \partial \Phi_2
\nonumber \\
&~~& - 2i D\Phi_1 D\Phi_3 \partial \Phi_3 + 2 D\Phi_2 \partial \Phi_2 D \Phi_3
\ena
Setting $\b^2=0$ the classical current in eq.(\ref{100}) is recovered.

\vskip 10pt
\noindent
{\bf 2) ${\bf B(1,1)}$ Toda theory}

For the case of the $B(1,1)$ superalgebra a convenient choice for
the two simple roots is (cfr. eq.(A.19))
\bea
\vec{\a}_1 &=& \vec{\varepsilon}_1 - \vec{\delta}_1 = (1,-i) \nonumber \\
\vec{\a}_2 &=& \vec{\delta}_1 = (0,i)
\label{root2}
\ena
The corresponding supersymmetric Toda action is then given by
\EQ
S = \frac{1}{\b^2} \int d^2 z d^2 \theta \left[ D\Phi_1 \bar{D} \Phi_1 + D
\Phi_2 \bar{D} \Phi_2 + e^{\Phi_1-i\Phi_2} + 2e^{i\Phi_2} \right]
\label{15}
\EN
The classical $W$--algebra is obtained from the Miura
operator in eq.(\ref{miura2}) setting $n=1$. Explicitly we have
\EQ
(D - D\Phi_1)(D -iD\Phi_2)D(D + iD\Phi_2)(D + D\Phi_1) =
D^5 + \sum_{i=0}^2 W^{(2-\frac{i}{2})}~D^i
\EN
where
\bea
W^{(1)} &=&
-i D \Phi_1 \pa \Phi_1 -i D \Phi_2 \pa \Phi_2 + D\pa \Phi_2 +2i D \pa \Phi_1
\nonumber \\
W^{(\frac{3}{2})} &=&
- i \pa^2 \Phi_2 +i D \Phi_1 D\pa \Phi_1 - (\pa \Phi_2)^2 -i D \Phi_2
D\pa \Phi_2 \nonumber \\
{}~~~~&~& +2D \Phi_1 D\pa \Phi_2  -2iD \Phi_1 D \Phi_2  \pa \Phi_2
\nonumber \\
W^{(2)} &=&
-D \pa^2\Phi_1 +D\Phi_1 \pa^2\Phi_1-i D\Phi_1 \pa^2\Phi_2 +
iD \pa \Phi_2 \pa\Phi_1
-(\pa \Phi_2)^2  D \Phi_1 \nonumber \\
{}~~~~&~&+ \pa \Phi_2 D \Phi_2 \pa \Phi_1 -i D \Phi_2
D \pa \Phi_2 D \Phi_1 \nonumber \\
&=& \frac{1}{2} D(W^{(\frac{3}{2})} +DW^{(1)})
\label{21}
\ena
We note that $W^{(1)}$ gives again the stress-energy tensor
\EQ
W^{(1)} = -iD\vec{\Phi} \cdot \pa \vec{\Phi} + 2i \vec{\rho} \cdot
D \pa \vec{\Phi}
\EN
where $\vec{\rho}$ is the Weyl vector (see eq.(A.16) in the Appendix).

In Ref.\cite{b14} we have shown that the holomorphic currents
in eq.(\ref{21}) maintain their form at the quantum level too,
albeit the coefficients of the various terms acquire a coupling constant
dependence. We report here the results and refer for
 details of the calculation to Ref.\cite{b14}.
The quantum holomorphic stress-energy tensor is given by
\EQ
W^{(1)}= -i D\Phi_1 \pa \Phi_1- i D\Phi_2 \pa \Phi_2  +(1- \frac{\b^2}{4 \pi})
D\pa \Phi_2 + 2i(1- \frac{\b^2}{8 \pi})D \pa \Phi_1
\label{17}
\EN
We note that the stress--energy tensor is
renormalized according to eq.(\ref{16}): in this case
one of the fermionic roots has nonvanishing norm.
For the spin--$\frac{3}{2}$ current one obtains the following
result
\bea
W^{(\frac{3}{2})}&=& -i (1-\frac{\b^2}{4\pi}) \pa^2 \Phi_2
+i(1-\frac{\b^2 }{2\pi}) D \Phi_1 D \pa \Phi_1 - (\pa \Phi_2)^2
-i D \Phi_2 D \pa \Phi_2 \nonumber \\
&~&+2(1-\frac{\b^2}{4\pi}) D\Phi_1 D \pa \Phi_2 -2i D \Phi_1 D \Phi_2
\pa \Phi_2
\label{18}
\ena
The  quantum version for the spin--$2$ current can be obtained from eqs.
(\ref{17}) and (\ref{18})
since  $W^{(2)}$ is linearly dependent on  $W^{(1)}$ and $W^{(\frac{3}{2})}$.

\vskip 10pt
\noindent
{\bf 3) ${\bf A(1,1)}$ Toda theory}

For the rank--$2$ superalgebra $A(1,1)$
an explicit realization of the fermionic simple roots is (cfr. eq.(A.24))
\bea
\vec{\a}_1 &=& \vec{\varepsilon}_1 - \vec{\delta}_1 = (1,-i) \nonumber \\
\vec\a_2 &=& \vec{\varepsilon}_1 + \vec{\delta}_1 = (1,i) \nonumber \\
\vec{\a}_3 &=& -\vec{ \a}_1
\label{root3}
\ena
Using this basis, the Toda action takes the form
\EQ
S = \frac{1}{\b^2} \int d^2z d^2\theta \left[ D\Phi_1 \bar{D} \Phi_1 + D \Phi_2
\bar{D} \Phi_2 + e^{\Phi_1 -i\Phi_2} + e^{\Phi_1+i\Phi_2} +
e^{-\Phi_1+i\Phi_2} \right]
\label{19}
\EN
The Miura operator in eq.(\ref{miura3}) for $n=1$ becomes
\EQ
D(D-D\Phi_1 +iD\Phi_2)(D+2iD\Phi_2)(D+D\Phi_1+iD\Phi_2)=D^4 + \sum_{i=0}^2
W^{(\frac{3}{2}-\frac{i}{2})}~D^i
\EN
where the spin--$s$ classical currents are
\bea
W^{(\frac12)} &=& -2\pa \Phi_2 + 2iD\Phi_2 D\Phi_1 \nonumber \\
W^{(1)} &=& iD\Phi_1 \pa \Phi_1 + iD\Phi_2 \pa \Phi_2 - iD\pa \Phi_1 - D\pa
\Phi_2 + i D(D\Phi_2 D\Phi_1) \nonumber \\
W^{(\frac{3}{2})} &=& -D(W^{(1)} - DW^{(\frac12)})
\label{20}
\ena
The classical stress--energy tensor is given by the holomorphic current
$W^{(1)}$, whereas $W^{(\frac12)}$ and $W^{(\frac{3}{2})}$ simply satisfy
the conservation equations
\EQ
\bar{D} W^{(\frac12)} = -2 D e^{-\Phi_1 +i\Phi_2} \qquad \qquad ~~
\bar{D} W^{(\frac{3}{2})} = -2i D\pa e^{-\Phi_1 +i\Phi_2}
\EN

The expressions in eq.(\ref{20}) give also the correct currents for the
quantum system since in this case no anomalies appear in the study of the
quantum conservation laws in eq.({\ref{9}).

\sect{Supercurrents of affine supersymmetric Toda theories}

We consider now the untwisted affine supersymmetric Toda theories described
by the action in eq.(\ref{1}). These models are not conformally invariant due
to the presence of a mass scale. The
conservation equation of the stress--energy tensor is then
modified by the appearance
of a nonvanishing trace $\bar{T}$
\EQ
\bar{D} T + D \bar{T} = 0
\EN
where $T$ can still be written in the form
\EQ
T= -iD\vec{\Phi} \cdot \pa \vec{\Phi} + 2i\vec{\rho} \cdot D \pa \vec{\Phi}
\EN

In order to exhibit the integrability of these theories
we have to prove that they possess higher--spin
currents $J^{(s)}$, $s>1$, satisfying the
conservation law
\EQ
\bar{D} J^{(s)} + D \bar{J}^{(s)} = 0
\label{22}
\EN
which reduces to eq.(4.1) for $s=1$.
As for the bosonic case \cite{b11}, the classical integrability
should be guaranteed by the Lax equations (\ref{2}) \cite{b15}. However,
for affine Toda theories based on Lie
superalgebras a general procedure
is still lacking.

For the simplest cases of the $D^{(1)}(2,1)$, $B^{(1)}(1,1)$ and
$A^{(1)}(1,1)$ theories we have established the existence of higher--spin
conserved currents at the classical and quantum level, by using the
approach described in the previous section. We compute
\EQ
\bar{D}_Z \left \langle J^{(s)}(Z,\bar{Z}) \right \rangle \equiv \bar{D}_Z
\left \langle J^{(s)}(Z,\bar{Z})
\exp \left(\frac{i}{\b^2}
\int d^2w d^2 \theta' {\cal L}_{int} \right) \right \rangle _0
\label{23}
\EN
and search for anomalous {\em local}
terms which are not expressible
as $D$-derivatives of some appropriate $\bar{J}$.
Any current of the form
$J^{(s)}= D {\cal J}^{(s-\frac{1}{2})}$ is not relevant since it trivially
satisfies eq.(\ref{22}). Furthermore, since we are not planning to determine
the actual form of $\bar{J}$, in the course of the calculation
we drop total $D$--derivatives and freely
integrate by parts on $z,\theta$.
We concentrate here on the explicit construction of the quantum
$J^{(s)}$ conserved currents
up to total $D$--derivatives.

\vskip 10pt
\noindent
{\bf 1) ${\bf D^{(1)}(2,1)}$ Toda theory}

The untwisted affine extension of the $D(2,1)$ theory is obtained by the
addition to the set of simple roots in eq.(\ref{root1}) of the corresponding
fermionic lowest root $\vec{\a}_0 =
-(\vec{\a}_1 + \vec{\a}_2 + \vec{\a}_3) = (-1,0,-i)$.
The supersymmetric action is
\EQ
S= \frac{1}{\b^2} \int d^2 z d^2 \theta \left[ D \vec{\Phi} \cdot \bar{D}
\vec{\Phi} + e^{\Phi_1 -i\Phi_3} + e^{-\Phi_2 +i\Phi_3} + e^{\Phi_2
+i\Phi_3} + e^{-\Phi_1-i\Phi_3} \right]
\label{24}
\EN
The theory is still renormalizable; as for the conformal case the only
ultraviolet divergences arise  from tadpole--type diagrams
and they can be cancelled by
normal ordering the exponentials in the Lagrangian.

The action in eq.(\ref{24})
is invariant under the following discrete symmetries: $\Phi_1 \rightarrow
-\Phi_1$, $\Phi_2 \rightarrow -\Phi_2$ and $(\Phi_1,\Phi_2,\Phi_3) \rightarrow
(\Phi_2,\Phi_1,-\Phi_3)$. Therefore we look for higher--spin currents
with a definite parity under these symmetries. An explicit calculation
reveals that the theory does not possess conserved currents with spin
$s=\frac{3}{2}, 2, \frac{5}{2}$. In particular, the $W^{(\frac{3}{2})}$
current present in the conformal case does not survive the affinization
since it does not respect the above symmetries.

The first non trivial conserved current appears at level $s=3$.
To find its expression we write
the most general linear combination of terms with seven
$D$--derivatives acting on the superfields $\Phi_1,\Phi_2,\Phi_3$,
{\em even} under the symmetries of the Lagrangian. Up to total
$D$--derivative contributions, we have
\bea
J^{(3)} &=& a[D\pa \Phi_1 \pa^2 \Phi_1 +  D\pa \Phi_2 \pa^2 \Phi_2] + b D\pa
\Phi_3 \pa^2 \Phi_3 \nonumber \\
&~~& + c[D\Phi_1 \pa \Phi_1 \pa^2 \Phi_3 - D\Phi_2 \pa \Phi_2 \pa^2 \Phi_3]
+ d [(\pa \Phi_1)^2 D\pa \Phi_3 - (\pa \Phi_2)^2 D\pa \Phi_3] \nonumber \\
&~~& + e [(\pa \Phi_1)^3 D\Phi_1 + (\pa \Phi_2)^3 D\Phi_2] + f (\pa \Phi_3)^3
D\Phi_3 \nonumber \\
&~~& + g[ D\Phi_1 \pa \Phi_1 (\pa \Phi_3)^2 +  D\Phi_2 \pa \Phi_2 (\pa
\Phi_3)^2] \nonumber \\
&~~& + h[ D\Phi_1 D\pa \Phi_1 D \Phi_3 \pa \Phi_3 +  D\Phi_2 D\pa \Phi_2 D
\Phi_3 \pa \Phi_3] \nonumber \\
&~~& + j[ (\pa \Phi_1)^2 \pa \Phi_3 D\Phi_3 + (\pa \Phi_2)^2 \pa \Phi_3
D\Phi_3] \nonumber \\
&~~& + (k+in) D\Phi_1 \pa \Phi_1 (\pa \Phi_2)^2 + n D\Phi_1 D\pa \Phi_1
D\Phi_2 \pa \Phi_2 + k (\pa \Phi_1)^2 \pa \Phi_2 D \Phi_2 \nonumber \\
&~~& + q D\Phi_1 \pa \Phi_1 D\Phi_2 \pa \Phi_2 D\Phi_3
\label{25}
\ena
We insert $J^{(3)}$ in eq.(4.4) and perform the Wick
contractions with the interaction Lagrangian. We can exploit the symmetries
of the action and the current, so that we need consider only contractions
with one exponential.
Using identities valid up to integration by parts
and dropping total $D$--derivatives, the resulting local contributions
can be arranged as follows ($\a \equiv \frac{\b^2}{2\pi}$):
\bea
&& \bar{D} \left \langle J^{(3)}
\left( \frac{i}{\b^2} \int d^2w d^2 \theta' e^{\Phi_1-i\Phi_3} \right)
\right \rangle \nonumber \\
&& \leadsto \left(
\left[ ic+id-\frac{\a}{2}g -\frac{\a}{2}j+\left(1+\frac{\a}{2}\right)(k+in)
+\left( 2+\frac{\a}{2}\right)k +in \right. \right. \nonumber \\
&&~~~~~~~~~~~~~~ \left. \left.  -\left(1+\frac{\a}{2}\right)q \right]
\pa \Phi_2 \pa^2 \Phi_2 \right. \nonumber \\
&& \left.
+\left[ -\frac12 c -\frac{i\a}{4}g -\frac{\a}{4} h +i\left(1+\frac{\a}{4}
\right)k -\frac12 n -\frac{i}{2} \left(1+\frac{\a}{2}\right)q  \right]
D\Phi_2 D\pa^2 \Phi_2 \right. \nonumber \\
&& \left.
+\left[ ig+ij-i(k+in) -ik+\frac{i}{2}q\right]
(\pa \Phi_2)^2 \pa \Phi_3
+ \left[-g+ih+k-\frac12 q \right] D\Phi_2 D\pa \Phi_2 \pa \Phi_3
\right. \nonumber \\
&& \left.
+\left[-g +\frac{\a}{4} q \right] D\Phi_2 \pa \Phi_2 D\pa \Phi_3
+\left[\frac{i}{2} h +j-(k+in) +\frac{i}{2} n \right]
\pa \Phi_2 D \pa \Phi_2 D \Phi_3 \right.
\nonumber \\
&&\left.
+\left[ -\frac{i}{2} h +\frac{i}{2} n \right] D\Phi_2 \pa^2 \Phi_2 D\Phi_3
+\left[n -ik +\frac{i}{2} \left(1+\frac{\a}{2} \right)q \right]
D\Phi_2 \pa \Phi_2 D\pa \Phi_1  \right. \nonumber \\
&&\left.
+\left[ -a-\frac{i}{2} \left(1+\frac{\a}{2} \right)c-i\left(1+\frac{\a}{4}
\right)d + \left(1+\frac{9\a}{4} +\frac{\a^2}{4} \right) e \right. \right.
\nonumber \\
&&~~~~~~~~~~~~~~ \left. \left.
-\left(\frac{\a}{4}+\frac{\a^2}{8} \right) g -\frac{i\a}{4} h - \left(
\frac{\a}{2} + \frac{\a^2}{8} \right) j \right] \pa \Phi_1 \pa^2 \Phi_1
\right. \nonumber \\
&& \left.
+\left[ ia -\frac12 \left(-1+ \frac{\a}{2} \right) c - \frac{\a}{4} d + i
\left( 2+ \frac{3\a}{4} -\frac{\a^2}{4} \right) e - i\left(\frac{5\a}{4}
-\frac{\a^2}{8} \right)g \right. \right. \nonumber \\
&&~~~~~~~~~~~~~~ \left. \left.
+ \frac12 \left(1-\frac{\a}{2} \right) h -
i\left(1+\a -\frac{\a^2}{8} \right)j \right] \pa \Phi_3 \pa^2 \Phi_1 \right.
\nonumber \\
&& \left.
+\left[ib -\left(1+\frac{\a}{4} \right) c - \left(1+\frac{\a}{4} \right) d
+2ie + i\left( \frac{3\a}{4} + \frac{\a^2}{4} \right) f
-i\left( 1+\a+\frac{\a^2}{8} \right) g \right. \right. \nonumber \\
&&~~~~~~~~~~~~~~ \left. \left.
+ \frac12 h -i \left(2+\frac{5\a}{4} +
\frac{\a^2}{8} \right) j \right] \pa \Phi_1 \pa^2 \Phi_3 \right. \nonumber
\\
&& \left.
+\left[ b+\frac{i\a}{4} c + \frac{i\a}{4} d + \left( 1-\frac{9\a}{4}
+\frac{\a^2}{4} \right) f +\left(\frac{\a}{2}-\frac{\a^2}{8} \right) g +
\left( \frac{\a}{4}-\frac{\a^2}{8} \right) j \right] \pa \Phi_3 \pa^2
\Phi_3 \right. \nonumber \\
&& \left.
+\left[ -\frac{i}{2} c -\frac{3\a}{4} f + \left(1+\frac{\a}{2} \right) g
+\frac{i}{2} \left(1- \frac{\a}{4} \right) h + \left(1+ \frac{\a}{4}
\right) j \right] D\Phi_1 D\pa^2 \Phi_3 \right. \nonumber \\
&& \left.
+\left[ \frac{i}{2} c - \left( 3+\frac{3\a}{4} \right) e + \frac{\a}{4} g
+\frac{i\a}{8} h + \left(1+ \frac{\a}{2} \right)j \right] D\Phi_3 D\pa^2
\Phi_1 \right. \nonumber \\
&& \left.
+\left[ id +3e +\frac{\a}{2} g +\frac{i}{2} (1-\a) h - \left(1+\frac{\a}{2}
\right) j \right] D\pa \Phi_1 D\pa \Phi_3 \right. \nonumber \\
&& \left.
+\left[ -3ie +ig -\frac12 h +2ij \right]  D\pa \Phi_1
D\Phi_3 \pa \Phi_3 + \left[ 2e+2f-2g-2j \right] (\pa \Phi_3)^2 \pa \Phi_1
\right. \nonumber \\
&& \left.
+\left[ -3if +2ig +\frac12 h +ij \right] \pa \Phi_3 D\pa \Phi_3
D\Phi_1  \right) e^{\Phi_1-i\Phi_3}
\label{26}
\ena
Since the various terms in the r.h.s. of
eq.(\ref{26}) are all independent up to total
$D$--derivatives, the $J^{(3)}$ current is conserved only if
the coefficients separately vanish. This condition gives
a system of equations which determine the unknowns $a,b, \cdots, q$ uniquely,
up to an overall factor. The $J^{(3)}$ quantum
conserved current is
\bea
J^{(3)} &=& -i\left(4 + \frac{3\b^2}{8\pi} -
\frac{35\b^4}{64\pi^2} \right)\left[ D\pa \Phi_1 \pa^2 \Phi_1 +
D\pa \Phi_2 \pa^2 \Phi_2 \right] \nonumber \\
&~~& -i\left(1 -\frac{3\b^2}{4\pi} - \frac{35\b^4}{64\pi^2} \right)
D\pa \Phi_3 \pa^2 \Phi_3
+i\left(1+\frac{3\b^2}{8\pi} \right)
(\pa \Phi_3)^3 D\Phi_3
\nonumber \\
&~~&-6 \left(1+\frac{3\b^2}{8\pi} \right)\left[ D\Phi_1 \pa \Phi_1 \pa^2 \Phi_3
 - D\Phi_2 \pa \Phi_2 \pa^2 \Phi_3 \right]
\nonumber \\
&~~&+ 6\left(1+\frac{\b^2}{4\pi}\right)\left[ (\pa \Phi_1)^2 D\pa \Phi_3
-(\pa \Phi_2)^2 D\pa \Phi_3 \right] \nonumber \\
&~~& -i\left(1-\frac{3\b^2}{8\pi} \right) \left[
(\pa \Phi_1)^3 D\Phi_1  + (\pa \Phi_2)^3 D\Phi_2 \right]
\nonumber \\
&~~& -i \frac{3\b^2}{2\pi}\left[  D\Phi_1 \pa \Phi_1 (\pa \Phi_3)^2 +
D\Phi_2 \pa \Phi_2 (\pa \Phi_3)^2 \right] \nonumber \\
&~~& -6 \left(1+\frac{5\b^2}{8\pi} \right)
\left[ D\Phi_1 D\pa \Phi_1 D \Phi_3 \pa \Phi_3
+ D\Phi_2 D\pa \Phi_2 D \Phi_3 \pa \Phi_3 \right]
\nonumber \\
&~~& +i\frac{9\b^2}{4\pi}\left[ (\pa \Phi_1)^2 \pa \Phi_3 D\Phi_3
+(\pa \Phi_2)^2 \pa \Phi_3 D\Phi_3 \right] \nonumber \\
&~~& -6i \left(1 + \frac{\b^2}{4\pi} \right)
D\Phi_1 \pa \Phi_1 (\pa \Phi_2)^2 -6
\left(1+\frac{5\b^2}{8\pi} \right) D\Phi_1 D\pa \Phi_1 D\Phi_2 \pa \Phi_2
\nonumber \\
&~~& +i\frac{9\b^2}{4\pi} (\pa \Phi_1)^2 \pa \Phi_2 D \Phi_2
-12i D\Phi_1 \pa \Phi_1 D\Phi_2 \pa \Phi_2 D\Phi_3
\ena
We note that the classical current obtained in the limit $\b^2
\rightarrow 0$ can be written, up to total $D$--derivative terms,
as a function of the $W$ and $F^{(1)}$ currents in eqs.(\ref{100},\ref{101})
\EQ
J^{(3)} = W^{(1)} D W^{(1)} + 2 W^{(1)} W^{(\frac{3}{2})} +4F^{(1)} D F^{(1)}
\label{27}
\EN
In the conformal case the three terms are separately
holomorphic currents, whereas in the
perturbed theory only the linear combination in eq.(\ref{27})
satisfies the classical conservation law. At the quantum level,
since $J^{(3)}$ is a composite
operator we could not simply insert the quantum expressions
of the $W^{(1)}$ and $W^{(\frac{3}{2})}$ currents in eq.(\ref{27}).
The lengthy calculation in eq.(\ref{26}) was
needed in order to obtain the complete renormalized form of $J^{(3)}$.

\vskip 10pt
\noindent
{\bf 2) ${\bf B^{(1)}(1,1)}$ Toda theory}

With the choice of simple roots in eq.(\ref{root2}), the lowest root of the
$B^{(1)}(1,1)$ model is
given by $\vec{\a}_0 = -(\vec{\a}_1 + 2\vec{\a}_2) = (-1,-i)$. The
affine action is
\EQ
S = \frac{1}{\b^2} \int d^2z d^2 \theta \left[ D\vec{\Phi} \cdot \bar{D}
\vec{\Phi} + e^{\Phi_1-i\Phi_2} + 2e^{i\Phi_2} + e^{-\Phi_1-i\Phi_2}
\right]
\label{30}
\EN
This system does not have nontrivial
conserved currents of spin $s= \frac{3}{2}$ or $s=2$. In particular
the $W^{(\frac{3}{2})}$ current is not conserved in the affine case
because it does not respect the discrete symmetry
$\Phi_1 \rightarrow - \Phi_1$ of the action.

The first higher--spin current
appears at $s=3$. Its construction is performed following the
same steps as for the $D^{(1)}(2,1)$ case.
We consider up to total $D$--derivative terms the most
general expression, {\em even} in $\Phi_1$
\bea
J^{(3)} &=& a D\pa \Phi_1 \pa^2 \Phi_1 +b D \pa \Phi_2 \pa^2 \Phi_2 +
c D \Phi_1 \pa \Phi_1 \pa^2 \Phi_2
+ d(\pa \Phi_1)^2 D \pa \Phi_2 \nonumber \\
&+& e (\pa \Phi_1 )^3 D \Phi_1 +
f ( \pa \Phi_2)^3 D \Phi_2 + g D \Phi_1 \pa \Phi_1 (\pa \Phi_2)^2
\nonumber \\
&+& h D\Phi_1 D \pa \Phi_1 D \Phi_2 \pa \Phi_2 +
k (\pa \Phi_1)^2 \pa \Phi_2 D \Phi_2
\ena
and compute all the local contributions which arise from Wick contracting with
the interaction Lagrangian. The anomaly cancellation requirement
leads to a set of equations for
$a,b,\dots ,k$ which can be solved {\em nontrivially}.
Details of the calculation are contained in Ref.\cite{b14}.
We quote here the result for the quantum $J^{(3)}$ current
\bea
J^{(3)} &=& i \left(-4 + \frac{13}{8\pi} \b^2 + \frac{7}{32\pi^2}
{\b}^4 - \frac{{\b}^6}{64\pi^3}  \right) D \partial \Phi_1 \partial^2
\Phi_1  \nonumber \\
&-& i \left(1 - \frac{7}{8\pi} \b^2 - \frac{7}{32\pi^2} \b^4 +
\frac{\b^6}{64\pi^3} \right) D \partial \Phi_2 \partial^2 \Phi_2  \nonumber \\
&+& \left(-6 + \frac{3}{2\pi}\b^2 \right)D \Phi_1 \partial \Phi_1 \partial^2
\Phi_2 + \left(6 - \frac{9}{4\pi} \b^2 + \frac{3}{16\pi^2}\b^4 \right)
(\partial \Phi_1)^2 D \partial \Phi_2  \nonumber \\
&-& i \left(1 - \frac{\b^2}{2\pi}
\right) (\partial \Phi_1)^3 D \Phi_1 +
i \left(1 + \frac{\b^2}{4\pi} \right) (\partial \Phi_2)^3 D \Phi_2
\nonumber \\
&-& 6 D \Phi_1 D \partial \Phi_1 D \Phi_2 \partial \Phi_2  +
\frac{3i}{4\pi}\b^2
(\partial \Phi_1)^2 \partial \Phi_2 D \Phi_2
\ena
The classical current is obtained by setting
$\b^2 = 0$. At the classical level
$J^{(3)}$ is equivalent,
up to total $D$--derivative terms, to the classical current
$W^{(1)}DW^{(1)}+2 W^{(1)}
W^{(\frac{3}{2})}$.

\vskip 10pt
\noindent
{\bf 3) ${\bf A^{(1)}(1,1)}$ Toda theory}

A particular realization of the simple roots for the $A(1,1)$ superalgebra
is the one in eq.(\ref{root3}). The corresponding fermionic lowest
root is then given by $\vec{\a}_0 = -\vec{\a}_2 = (-1,-i)$.
The $A^{(1)}(1,1)$ affine Toda action is
\EQ
S = \frac{1}{\b^2} \int d^2z d^2 \theta \left[ D \vec{\Phi} \cdot \bar{D}
\vec{\Phi} + e^{\Phi_1-i\Phi_2} + e^{\Phi_1+i\Phi_2} + e^{-\Phi_1+i\Phi_2}
+ e^{-\Phi_1-i\Phi_2} \right]
\label{31}
\EN
It is symmetric under $\Phi_1 \rightarrow
-\Phi_1$ and $\Phi_2 \rightarrow -\Phi_2$. The first
nontrivial conserved current appears again at $s=3$.
The most general linear combination of
terms with seven $D$--derivatives, {\em even} under the above mentioned
discrete symmetries is
\bea
J^{(3)} &=& a D\pa \Phi_1 \pa^2 \Phi_1 + b D\pa \Phi_2 \pa^2 \Phi_2 + c
D\Phi_1 (\pa \Phi_1)^3 + d D\Phi_2 (\pa \Phi_2)^3 \nonumber \\
&~~& + e (\pa \Phi_1)^2 D\Phi_2 \pa \Phi_2 + f (\pa \Phi_2)^2 D\Phi_1 \pa
\Phi_1 \nonumber \\
&~~& + g D\Phi_1 D\pa \Phi_1 D\Phi_2 \pa \Phi_2
\ena
We find ($\a \equiv \frac{\b^2}{2 \pi}$)
\bea
&& \bar{D} \left \langle J^{(3)}
\left( \frac{i}{\b^2} \int d^2w d^2 \theta' e^{\Phi_1+i\Phi_2} \right)
\right \rangle \nonumber \\
&& \leadsto \left(
\left[ -a + (1+\frac{9\a}{4}+ \frac{\a^2}{4})c-(\frac{\a}{2}+\frac{\a^2}{8})e
-(\frac{\a}{4}+\frac{\a^2}{8})f -\frac{i\a}{4} g \right] \pa \Phi_1
\pa^2 \Phi_1 \right. \nonumber \\
&& \left.
+\left[ b+(1-\frac{9\a}{4}+\frac{\a^2}{4})d +(\frac{\a}{4}-\frac{\a^2}{8})e +
(\frac{\a}{2}-\frac{\a^2}{8})f \right] \pa\Phi_2 \pa^2 \Phi_2
\right. \nonumber \\
&& \left.
+\left[ -ia -i(2+\frac{3\a}{4}-\frac{\a^2}{4})c + i(1+\a -\frac{\a^2}{8})e
+i(\frac{5\a}{4}-\frac{\a^2}{8})f \right. \right. \nonumber \\
&&~~~~~~~~~~~~~~ \left. \left. +(-\frac12 +\frac{\a}{4})g \right]
\pa \Phi_2 \pa^2 \Phi_1 \right.
\nonumber \\
&&\left.
+\left[ -ib -2ic -i(\frac{3\a}{4}+\frac{\a^2}{4})d +i(2+\frac{5\a}{4}
+\frac{\a^2}{8})e +i(1+\a +\frac{\a^2}{8})f -\frac12 g \right]
\pa \Phi_1 \pa^2 \Phi_2 \right. \nonumber \\
&& \left.
+\left[ -3c+(1+\frac{\a}{2})e -\frac{\a}{2} f -\frac{i}{2}(1-\a)g \right]
D\pa\Phi_1 D\pa \Phi_2 \right. \nonumber \\
&& \left.
+\left[ -3(1+\frac{\a}{4})c +(1+\frac{\a}{2})e + \frac{\a}{4}f
+\frac{i\a}{8} g \right] D\pa^2 \Phi_1 D\Phi_2 \right. \nonumber \\
&& \left.
+\left[ \frac{3\a}{4}d - (1+\frac{\a}{4})e - (1+\frac{\a}{2}) f +
\frac{i}{2}(-1+\frac{\a}{4}) g \right] D\Phi_1 D\pa^2 \Phi_2 \right.
\nonumber \\
&& \left.
+\left[ 3id -ie -2if -\frac12 g \right] D\Phi_1 \pa\Phi_2 D\pa \Phi_2 +
\left[2c +2d-2e-2f \right]
\pa \Phi_1 (\pa \Phi_2)^2 \right. \nonumber \\
&& \left.
+\left[ -3ic+2ie+if-\frac12 g \right] D\pa \Phi_1 D\Phi_2 \pa \Phi_2
\right) e^{\Phi_1+i\Phi_2}
\label{32}
\ena
where the result has been expressed in terms of an independent set. In this
case
too the contractions with the other exponentials need not be considered
on the basis of the symmetries of the Lagrangian.
The quantum current is determined, up to
an overall normalization factor, by imposing that the r.h.s. vanishes. We find
\bea
J^{(3)} &=& -i\left(1 - \frac{11\b^4}{16\pi^2}\right)
[D\pa \Phi_1 \pa^2 \Phi_1 + D\pa \Phi_2 \pa^2 \Phi_2] -i
\left(1-\frac{3\b^2}{4\pi} \right)
D\Phi_1 (\pa \Phi_1)^3 \nonumber \\
&~~& +i \left( 1 +\frac{3\b^2}{4\pi} \right) D\Phi_2 (\pa \Phi_2)^3
-i \left( 3-\frac{9\b^2}{4\pi} \right) (\pa \Phi_1)^2 D\Phi_2 \pa \Phi_2
\nonumber \\
&~~& +i\left( 3-\frac{3\b^2}{4\pi} \right) (\pa \Phi_2)^2 D\Phi_1 \pa
\Phi_1
- \frac{3\b^2}{\pi} D\Phi_1 D\pa \Phi_1 D\Phi_2 \pa \Phi_2
\ena
Setting $\b^2=0$ we obtain the classical $J^{(3)}$ current which is expressible
as a linear combination of the $W$--currents
\EQ
J^{(3)} = W^{(1)} D W^{(1)} - iW^{(1)} \pa W^{(\frac12)} + \frac12
(W^{(\frac12)})^2 W^{(1)}
\EN

\sect{The mass spectrum}

In this section
we study the classical mass spectrum of the
fundamental particles of the affine Toda theories based on the
$D^{(1)}(n+1,n)$, $B^{(1)}(n,n)$ and $A^{(1)}(n,n)$ superalgebras,
and for the first simplest cases we compute one--loop corrections
to the masses.
{}From the general expression of the affine
supersymmetric Toda action in eq.(2.8),
expanding the potential up to third order, we obtain the mass matrix and the
three--point couplings
\bea
S &=& \frac{1}{\b^2} \int d^2z d^2\theta \left[ D\vec{\Phi} \cdot \bar{D}
\vec{\Phi} + \frac12 \sum_{i=0}^r q_i \a_i^a \a_i^b \Phi_a \Phi_b +
\frac{1}{3!} \sum_{i=0}^r q_i \a_i^a \a_i^b \a_i^c \Phi_a \Phi_b \Phi_c
+O(\Phi^4) \right]
\nonumber \\
&\equiv&
\frac{1}{\b^2} \int d^2z d^2 \theta \left[ \vec{\Phi} (\bar{D} D - M)
\vec{\Phi} + c^{abc} \Phi_a \Phi_b \Phi_c  +O(\Phi^4) \right]
\label{34}
\ena
The first order term is zero due to the definition of the Ka$\check{c}$
labels $q_i$ ($\sum_{i=0}^r q_i \vec{\a_i} = 0$).
The second order term contains the matrix
\EQ
M^{ab} \equiv - \frac12 \sum_{i=0}^r q_i \a_i^a \a_i^b
\EN
whose eigenvalues give the supersymmetric classical mass spectrum.
In terms of the mass eigenvalues $M_k$, the masses of the bosonic particles
are $m_k^2 = 2M_k^2$. With the choice of simple roots listed in the Appendix,
the mass matrix in eq.(5.2)
for the $D^{(1)}(n+1,n)$, $B^{(1)}(n,n)$ and $A^{(1)}(n,n)$ models is not
diagonal, so that in order to determine the mass spectrum one has to solve
the corresponding eigenvalue problem. We have not succeeded in doing this for
general $n$, but we
have studied a number of theories
and guessed the general expression of
the masses $m_k$. They are given by
\bea
D^{(1)}(n+1,n) &:& ~~~~~~~~ m_k^2 = 8 {\sin}^2 \frac{2\pi k}{h}
\qquad \qquad \qquad \qquad \qquad k=1,
\cdots , 2n-1  \nonumber \\
&~~& ~~~~~~~~ m_{2n}^2 = m_{2n+1}^2 = 2 \\
&~~&~~~~~~ \nonumber \\
B^{(1)}(n,n) &:& ~~~~~~~~ m_k^2 = 8 {\sin}^2 \frac{2\pi k}{h}
\qquad \qquad \qquad \qquad \qquad k=1, \cdots ,2n-1 \nonumber \\
&~~& ~~~~~~~~ m_{2n}^2 = 2 \\
&~~&~~~~~ \nonumber \\
A^{(1)}(n,n) &:& ~~~~~~~~ m_k^2 = m^2_{2n+1-k} = 8 {\sin}^2 \frac{2\pi k}{h}
\qquad \qquad ~~~ k=1, \cdots ,n
\ena
where $h$ is the Coxeter number of the superalgebras,
$h= 4n$ for $D^{(1)}(n+1,n)$ and $B^{(1)}(n,n)$, $h=2n+2$ for
$A^{(1)}(n,n)$.
We have checked our guess with the computer program {\em Mathematica} for
the cases $n=1, \cdots , 6$.

We note that the mass spectrum for the $B^{(1)}(n,n)$ theory is naturally
obtained from the spectrum of the $D^{(1)}(n+1,n)$ model by setting
$\Phi_{2n+1} \equiv 0$, which corresponds to the folding operation from
the $B^{(1)}(n,n)$ Dynkin diagram to the $D^{(1)}(n+1,n)$ one.

We stress that the masses of all these theories are real
despite the manifest nonhermiticity of the action in eq.(\ref{1}).

In the $D$--theories the $n$, $2n$ and $2n+1$ particles
are at threshold, due to the particular mass relations
$m_n=2m_{2n}=2m_{2n+1}$. In the same way we have $m_n=2m_{2n}$ for the
$B$--theories. As a result of this,
divergent contributions similar to the ones discussed in Ref.\cite{b16}, could
be produced in on--shell amplitudes. However, one can show that
wave--function renormalizations are sufficient to make these theories well
defined.

In order to compute one--loop mass corrections one needs evaluate on--shell
self--energy supergraphs. It is convenient to perform the calculation
using a basis of superfields in which the mass matrix $M$ is diagonal.
In this way we have massive propagators for each internal line of a
Feynman supergraph
\EQ
\langle \Phi_i(Z,\bar{Z}) \Phi_j(0,0) \rangle = -i \delta_{ij}
\b^2 \frac{(\bar{D}D +M_i)}{\Box +2M_i^2} \delta^{(2)}(\theta)
\EN
whereas external lines satisfy the on--shell condition $\bar{D}D\Phi_j
= M_j \Phi_j$. The one--loop contribution to the effective action
$(e^{iS} \rightarrow e^{i\G})$ from a generic self--energy diagram with
particle $i$ on the external lines, and $j,k$ internal, is of the form
\EQ
 \int d^2\theta ~\Phi_i (\bar{D}D +M_j +M_k)\Phi_i ~\S(p_i^2;m_j^2,m_k^2)
\EN
where
\EQ
\S(p_i^2;m_j^2,m_k^2) = \frac{1}{(2\pi)^2} \int \frac{d^2 k}{(k^2
-m_j^2)[(k-p_i)^2 -m_k^2]}
\label{39}
\EN

We have computed one--loop mass corrections for the $D^{(1)}(2,1)$ and
$D^{(1)}(3,2)$ theories, for the $B^{(1)}(1,1)$ and $B^{(1)}(2,2)$ theories
and finally for the $A^{(1)}(1,1)$ and $A^{(1)}(2,2)$ models.
As already mentioned, in the $D$ and $B$ cases threshold divergences
are produced at intermediate stages of the calculation but they all
cancel in the final answer. We have found that in the $D^{(1)}(n+1,n)$
and $A^{(1)}(n,n)$ theories with $n=1,2$, the classical
mass spectrum is stable under quantization,
since the couplings and the masses for these theories
are such that the on--shell
contributions from self--energy supergraphs always sum up to
zero. Absence of one--loop mass renormalization is verified also for the two
supermultiplets described by the $B^{(1)}(1,1)$ model.
At level $n=2$, while the masses of the two superfields at threshold,
i.e. the lightest and the
heaviest, are still not renormalized, nonvanishing mass contributions are
produced for the
the other two. In general we expect
that the classical mass ratios of the $B^{(1)}(n,n)$ models will be spoiled by
quantum corrections following a
pattern which is typical of bosonic nonsimply-laced Toda theories. We show this
explicitly on the $B^{(1)}(2,2)$ example.

The classical Lagrangian of the $B^{(1)}(2,2)$ theory is given by
\EQ
{\cal L} = D\vec{\Phi}\cdot \bar{D} \vec{\Phi}+e^{\Phi_1-i\Phi_3}+
2e^{i\Phi_3-\Phi_2}+2e^{\Phi_2-i\Phi_4}+2e^{i\Phi_4}+e^{-\Phi_1-i\Phi_3}
\EN
In this basis, which corresponds to the realization of roots as in
eq.(A.19,A.20) for $n=2$, the mass matrix is not diagonal. We obtain a
diagonal basis through the change of variables
\bea
&~& \Phi_1 \rightarrow \Phi_4 \qquad \qquad \qquad \qquad\qquad
\Phi_2 \rightarrow \frac{i \sqrt{\sqrt{2}-1}}{\sqrt{2}} \Phi_1
+\frac{ \sqrt{\sqrt{2}+1}}{\sqrt{2}} \Phi_3 \nonumber \\
&~&\Phi_3 \rightarrow \frac{ \sqrt{\sqrt{2}+1}}{2} \Phi_1
+ \frac{1}{\sqrt{2}} \Phi_2 -\frac{ i\sqrt{\sqrt{2}-1}}{2} \Phi_3 \nonumber\\
&~&\Phi_2 \rightarrow \frac{ \sqrt{\sqrt{2}+1}}{2} \Phi_1
- \frac{1}{\sqrt{2}} \Phi_2 -\frac{ i\sqrt{\sqrt{2}-1}}{2} \Phi_3
\ena
The relabeling of the fields has been chosen in such a way that the
corresponding masses are the ones in eq.(5.4); specifically we have
\EQ
m_1^2=4\qquad \qquad m_2^2=8 \qquad \qquad m_3^2=4 \qquad \qquad m_4^2=2
\EN
The three--point couplings in the new basis are
\bea
{\cal L}^{(3)}
&=& \frac{i}{\sqrt{2}} \Phi_2 \Phi_1^2 + \frac{i}{\sqrt{2}} \Phi_2
\Phi_3^2 + \sqrt{2} \Phi_1 \Phi_2 \Phi_3 \nonumber \\
&~~& - \frac{i}{\sqrt{2}} \Phi_2 \Phi_4^2 - \frac{i\sqrt{\sqrt{2} + 1}}{2}
\Phi_1 \Phi_4^2 - \frac{\sqrt{\sqrt{2}-1}}{2} \Phi_3 \Phi_4^2
\ena
One-loop mass corrections arise from the diagrams shown in Fig.2.

In terms of the momentum integrals $\S(p_i^2;m_j^2,m_k^2)$ in eq.({\ref{39}),
the contributions from the various supergraphs evaluated on--shell are

\vskip 10pt

\halign{\quad  $#$  \hskip 0.80in & $#$ \cr
(a):~ \frac{1}{\sqrt{2}} \S(m_1^2;m_4^2,m_4^2)~~~~~~~~~ &
(b):~ -4(1+\sqrt{2}) \S(m_1^2;m_1^2,m_2^2) \cr
& \cr
(c):~ 4\S(m_1^2;m_2^2,m_3^2)~~~~~~~~~ & ~~~~~~~~~~~~~\cr
& \cr
(d):~ -2(\sqrt{2}+1) \S(m_2^2;m_1^2,m_1^2) &
(e):~ 2(\sqrt{2}-1) \S(m_2^2;m_3^2,m_3^2)\cr
& \cr
(f):~ 0~~~~~~~~~~~~~~~~~~~~~~~~~~~~~~~~~~ &
(g):~ 4\S(m_2^2;m_1^2,m_3^2)~~~~~~~~~~~ \cr
& \cr
(h):~ 4(\sqrt{2}-1) \S(m_3^2;m_2^2,m_3^2)~~~ &
(i):~ - \frac{1}{\sqrt{2}} \S(m_3^2;m_4^2,m_4^2)~~~~~~~~~ \cr
& \cr
(l):~ 4 \S(m_3^2;m_1^2,m_2^2)~~~~~~~~~~~~ & \cr
& \cr
(m):~ 0~~~~~~~~~~~~~~~~~~~~~~~~~~~~~~~~~~ &
(n):~ \sqrt{2} \S(m_4^2;m_1^2,m_4^2)~~~~~~~~~~ \cr
& \cr
(o):~ -\sqrt{2} \S(m_4^2;m_3^2,m_4^2)~~~~~~~~~~~ & \cr}
\EQ
{}~~~~~~~
\EN
Using the equalities $\S(m_1^2;m_1^2,m_2^2)=\S(m_1^2;m_2^2,m_3^2)$ and
$\S(m_1^2;m_4^2,m_4^2)=4\S(m_1^2;m_1^2,m_2^2)$ we obtain the
one--loop mass corrections
\bea
&&\Delta M_1 = - \Delta M_3 = -i\sqrt{2}\b^2 \S(m_1^2;m_1^2,m_2^2)
= \frac{\b^2}{32 \sqrt{2}} \nonumber\\
&& \Delta M_2 = \Delta M_4 = 0
\ena
Therefore the classical mass ratios of the $B^{(1)}(2,2)$ theory are not
maintained.

The analysis of the mass spectrum we have presented is not complete since
the affine supersymmetric Toda theories admit {\em soliton} configurations.
At the classical level the presence of such solutions is signalled by the
periodicity of the potential under a shift of the superfields associated
to the imaginary components of the roots
\EQ
\vec{\Phi} \rightarrow \vec{\Phi} +2\pi \vec{\o}
\EN
with $\vec{\o}\cdot\vec{\a}_i= k$ and $k$ any integer. For the $D^{(1)}(n+1,n)$
and $B^{(1)}(n,n)$ theories supersymmetric
soliton solutions can be constructed in a
straightforward manner since setting to zero the superfields associated to the
real components of the roots amounts to consider $n$ decoupled supersine-Gordon
systems. In these cases one can then borrow all the results in the literature
\cite{b17} and obtain the complete solitonic spectrum.

For the $A^{(1)}(n,n)$ theories, when $n\geq 2$
the situation is more complicated:
looking at the explicit expressions of the roots in eq.(A.24,A.25) it is clear
that the interaction between the $n$ superfields associated to the
imaginary components of the roots is highly nontrivial. The general
construction
of soliton solutions is beyond the scope of this paper.

\sect{Conclusions}

We have considered Toda theories based on Lie superalgebras for which an
extended set of fermionic roots exist:
the corresponding field theories are supersymmetric. More precisely
we have given an explicit lagrangian realization in $N=1$ superspace
for the $D(n+1,n)$, $B(n,n)$ and $A(n,n)$ theories.
These systems, although nonunitary,
are {\em quantum} integrable. We have constructed the superspace
Miura operators and computed the exact renormalized expressions of the
currents which determine the W-symmetries of the conformal theories.
The existence of quantum conserved higher--spin currents has then
been established
for specific examples in the untwisted affine cases, namely we have proven
conservation to all--loop order of the spin--$3$ current for the
$D^{(1)}(2,1)$,
$B^{(1)}(1,1)$ and $A^{(1)}(1,1)$ theories.

The affine theories are massive
systems: we have computed one--loop corrections to the particle mass spectrum
and found results that parallel the situation for bosonic Toda theories
based on simply-laced
and nonsimply-laced Lie algebras.

All the calculations we have presented have been obtained taking advantage of
a full superspace approach. We emphasize that this formalism is particularly
advantageous in the study of the renormalization of the supercurrents, since
a component calculation would have been prohibitively complicated.

Having exhibited the existence of
higher--spin charges which commute with the supersymmetric hamiltonians,
we could proceed to the determination of the S matrices, which are now
guaranteed to be elastic and factorizable.
For supersymmetric Toda theories, as compared to {\em unitary}
Toda systems, things are not so straightforward:
the presence of solitonic configurations
alters the mass spectrum in a substantial way and the full
quantum group approach is then required for the construction of the complete
S-matrix \cite{b18}.

\newpage

\appendix \sect{Appendix}

In this appendix we list some of the basic properties, the Dynkin diagrams and
the set of fermionic simple roots for superalgebras \footnote{A comprehensive
introduction to superalgebras is contained in Ref.\cite{b19}}
which admit a fermionic untwisted affine extension.
We also give the generators and
the weights in the fundamental representation, which are needed to construct
the Miura transformation for the corresponding Toda theories.

We remind that, given a rank--$r$ superalgebra with a completely
fermionic set of simple roots,
its generators $h_a$, $e_i^{+}$, $e_j^{-}$ satisfy
\EQ
[h_a,h_b] = 0 \qquad \qquad [h_a,e_j^{\pm}] = \pm \a^a_j
e_j^{\pm} \qquad \qquad
\{e_i^{+},e_j^{-}\} = \delta_{ij} h_a \a^a_i
\label{C}
\EN
where $\{~,~\}$ indicates the anticommutation operator.
The simple roots $\a_j$ are linear functionals on
the Cartan subalgebra ${\cal H}$
and they can be realized explicitly as $r$--dimensional vectors
with components $\a_j^a \equiv  \a_j(h_a) $.
It is convenient to represent
the roots in terms of vectors $\vec{\varepsilon}_i$,
$\vec{\delta}_j$, defined by
\EQ
\vec{\varepsilon}_i \cdot \vec{\varepsilon}_j = \delta_{ij} \qquad \qquad
\vec{\delta}_i \cdot \vec{\delta}_j = - \delta_{ij} \qquad \qquad
\vec{\varepsilon}_i \cdot \vec{\delta}_j = 0
\EN

We will give explicit expressions of the generators of the various
superalgebras in the fundamental representation. To this end we introduce
a set of complex matrices ${\cal M}(p|q;\cal{C})$
of the form
\EQ
M =
\left( \begin{array}{cc}
A & B \\
C & D
\end{array}
\right)
\EN
where the dimensions of the submatrices $A$, $B$, $C$ and $D$ are respectively
$p \times p$, $p \times q$ $q \times p$ and $q \times q$. The matrix $M$ is
said to be {\em even} if $C=B=0$, {\em odd} if $A=D=0$.
The generators $\vec{h}$, $e_j^{\pm}$
are then expressed as $d \times d$ such matrices, being
$d=p+q$ the dimension of the fundamental representation of the
superalgebra, in terms of the basis
\EQ
(e_{i,j})_{kl} = \delta_{ik} \delta_{jl}
\EN

The weights $\l$ can be obtained through the following general construction.
Given a representation $\G$ for the superalgebra,
the weights $\l$ are defined by the relation
\EQ
\G(h) \psi(\l) = \l(h) \psi(\l)
\label{weights}
\EN
where $\G(h)$ is a diagonal matrix which realizes the $\G$--representation
of an element $h$ in the Cartan subalgebra and $\psi(\l)$ is a
set of common eigenvectors for the matrices $\G(h)$.
Since the weights are linear functionals on ${\cal H}$, they
can be expressed as linear
combinations of the simple roots
\EQ
\l = \sum_{j=1}^{r} \mu_j \a_j
\label{eigen}
\EN
where the set of coefficients $\mu_j$ can be determined by solving the system
obtained from the relation (\ref{eigen}) evaluated on each element of the
${\cal H}$ basis:
\bea
\l(h_1) &=& \sum_{j=1}^r \mu_j \a_j(h_1) \nonumber \\
\l(h_2) &=& \sum_{j=1}^r \mu_j \a_j(h_2) \nonumber \\
& \cdots & \nonumber \\
\l(h_r) &=& \sum_{j=1}^r \mu_j \a_j(h_r)
\ena
Remembering that $\a_j(h_a) = \a_j^a$ are the components of the simple
roots, we can express the coefficients $\mu_j$ as linear functions of
$\l(h_1) \cdots \l(h_r)$. Finally,
if we define $\l_j(h) \equiv (\G(h))_{jj}$,
the weights in eq.(\ref{eigen}) are given by
\bea
\l_1 &=& \sum_{j=1}^r \mu_j(\G(h_1)_{11} \cdots
\G(h_r)_{11})~\a_j
\nonumber \\
\l_2 &=& \sum_{j=1}^r \mu_j(\G(h_1)_{22} \cdots
\G(h_r)_{22})~\a_j
\nonumber \\
& \cdots & \nonumber \\
\l_s &=& \sum_{j=1}^r \mu_j(\G(h_1)_{ss} \cdots
\G(h_r)_{ss})~\a_j
\ena
where $s$ is the dimension of the $\G$--representation. In the following we
will
work in the fundamental representation.

We explicitly give results for the superalgebras $D(n+1,n)$, $B(n,n)$
and $A(n,n)$.
The Dynkin diagrams of the corresponding
untwisted affine extensions are represented in Fig.1.
The diagrams encode informations about the
lengths of the roots $\vec{\a}^2_i$ and their inner products, defined as
$\vec{\a}_i
\cdot \vec{\a}_j = \sum_{k=1}^r \a_i^k \a_j^k$ (no complex conjugation). The
Ka$\check{c}$ labels are also explicitly shown, with $q_0=1$ so that
$\sum_{i=0}^r q_i \vec{\a}_i =0$. Their sum defines the Coxeter number $h$
\EQ
h= \sum_{i=0}^r q_i        \label{coxeter}
\EN

\vskip 20pt
\noindent
1) $D(n+1,n)$, $n \geq 1$

The $D(n+1,n)$ superalgebra is a simple superalgebra with rank $2n+1$.
An explicit  realization is given in terms of matrices in ${\cal
M}(2n+2|2n;\cal{C})$ satisfying the condition
\EQ
M^{{\rm st}} K + (-1)^{deg(M)} KM = 0
\label{cond}
\EN
where $M^{\rm st}$ indicates the supertranspose of the matrix $M$ and
\EQ
K =
\left(\begin{array}{cccc}
0 & {\bf I}_{n+1} & 0 & 0 \\
{\bf I}_{n+1} & 0 & 0 & 0 \\
0 & 0 & 0 & {\bf I}_{n} \\
0 & 0 & -{\bf I}_{n} & 0
\end{array} \right)
\EN
This provides the fundamental representation of the algebra $D(n+1,n)$.

The completely fermionic extended Dynkin diagram is given in Fig.$1$. A
corresponding choice of positive simple roots and generators is
\bea
\vec{\a}_1 &=& \vec{\varepsilon}_1 - \vec{\delta}_1 ~\quad \qquad \qquad \qquad
e_1^+ = e_{1,2n+3} + e_{3n+3,n+2} \nonumber \\
\vec{\a}_2 &=& \vec{\delta}_1 - \vec{\varepsilon}_2 ~\quad \qquad \qquad \qquad
e_2^+ = e_{n+3,3n+3} - e_{2n+3,2} \nonumber \\
& \cdots & ~~\qquad \qquad \qquad \quad \qquad \qquad \cdots \nonumber \\
\vec{\a}_{2n-1} &=& \vec{\varepsilon}_{n} - \vec{\delta}_{n} \qquad \qquad
\qquad e_{2n-1}^+ =
e_{n,3n+2} + e_{4n+2,2n+1} \nonumber \\
\vec{\a}_{2n} &=& \vec{\delta}_{n} - \vec{\varepsilon}_{n+1}
\qquad \qquad \qquad e_{2n}^+ = e_{2n+2,4n+2} - e_{3n+2,n+1}  \nonumber \\
\vec{\a}_{2n+1} &=& \vec{\delta}_{n} + \vec{\varepsilon}_{n+1}
\quad \qquad \qquad e_{2n+1}^+ = e_{n+1,4n+2} - e_{3n+2,2n+2}
\ena
while the lowest root is given by
\EQ
\vec{\a}_0 = -\vec{\varepsilon}_1 - \vec{\delta}_1
\quad \qquad \qquad e_0^+ = e_{n+2,2n+3} + e_{3n+3,1}
\EN
For any generator $e_j^+ = e_{ab} \pm e_{cd}$ associated to a
positive simple root, the generator associated to the corresponding negative
root is $e_j^- = e_{ba}\mp e_{dc}$.
An explicit expression for the roots can be obtained if we choose the
$\vec{\varepsilon}$~'s and $\vec{\delta}$~'s
as $(2n+1)$--dimensional vectors of the form
\bea
(\vec{\varepsilon}_j)_k &=& \delta_{kj} \qquad \qquad \qquad \qquad ~~~
j=1, \cdots , n+1 \nonumber \\
(\vec{\delta}_j)_k &=& i \delta_{k,j+n+1} ~~\qquad \qquad \qquad j=1, \cdots ,n
\ena
A convenient basis for the Cartan subalgebra is the following:
\bea
h_a &=& e_{a,a} - e_{a+n+1,a+n+1} \qquad \qquad \qquad \qquad a=1,\cdots ,n+1
\nonumber \\
h_a &=& i(e_{a+n+1,a+n+1} - e_{a+2n+1,a+2n+1})   ~\qquad a=n+2,\cdots ,2n+1
\ena
The Weyl vector $\vec{\r}$ of the algebra can be obtained as half
the sum of the vectors $\vec{\nu}_j$ dual to the simple
roots ($\vec{\a}_i \cdot \vec{\nu}_j = \delta_{ij}$). For
this case we have
\EQ
\vec{\rho} = n \vec{\varepsilon}_1 - (n-\frac12)\vec{\delta}_1
+ (n-1)\vec{\varepsilon}_2 -
(n-\frac{3}{2}) \vec{\delta}_2
+ \cdots +
\vec{\varepsilon}_{n} - \frac12 \vec{\delta}_{n}
\label{weyl}
\EN

The weights of the fundamental representation can be computed by
using the general procedure described above. They are
\bea
\vec{\l}_j &=& - \vec{\l}_{4n+3-j}~~ =~~ \sum_{k=j}^{2n-1} \vec{\a}_k +
\frac12 \vec{\a}_{2n} +
\frac12 \vec{\a}_{2n+1} \qquad \qquad j=1, \cdots, 2n-1 \nonumber \\
\vec{\l}_{2n} &=& ~ -\vec{\l}_{2n+3} ~~~~ =~~  \frac12 \vec{\a}_{2n} +
\frac12 \vec{\a}_{2n+1} \nonumber \\
\vec{\l}_{2n+1} &=& ~~ -\vec{\l}_{2n+2}~~~ =~~ -\frac12 \vec{\a}_{2n} +
\frac12 \vec{\a}_{2n+1}
\ena
For the $D(n+1,n)$ superalgebra,
the ordering of the weights is not unique. We have chosen the above labeling
so that the Miura operator
can be written in terms of the weights in a manner
similar to the $B(n,n)$ and $A(n,n)$ cases
(cfr. eqs.(\ref{miura1}, \ref{miura2},\ref{miura3})).

\vskip 10pt
\noindent
2) $B(n,n)$, $n \geq 1$

This is a simple superalgebra with rank $2n$. The fundamental
representation is given in terms of matrices in ${\cal M}(2n+1|2n;\cal{C})$
satisfying the condition (\ref{cond}) where in this case $K$ is given by
\EQ
K = \left(\begin{array}{ccccc}
0 & {\bf I}_n & 0 & 0 & 0 \\
{\bf I}_n & 0 & 0 & 0 & 0 \\
0 & 0 & 1 & 0 & 0 \\
0 & 0 & 0 & 0 & {\bf I}_n \\
0 & 0 & 0 & -{\bf I}_n & 0
\end{array} \right)
\EN
The fermionic set of simple roots corresponding to the Dynkin diagram in
Fig.$1$ and the generators $e_j^+$ are
\bea
\vec{\a}_1 &=& \vec{\varepsilon}_1 - \vec{\delta}_1 ~~
\quad \qquad \qquad \qquad
e_1^+ = e_{1,2n+2} + e_{3n+2,n+1} \nonumber \\
\vec{\a}_2 &=& \vec{\delta}_1 - \vec{\varepsilon}_2 ~~
\quad \qquad \qquad \qquad
e_2^+ = e_{n+2,3n+2} - e_{2n+2,2} \nonumber \\
& \cdots & ~~~\qquad \quad \qquad \qquad \qquad \qquad \cdots \nonumber \\
\vec{\a}_{2n-2} &=& \vec{\delta}_{n-1} - \vec{\varepsilon}_n
{}~\quad \qquad \qquad e_{2n-2}^+ = e_{2n,4n} - e_{3n,n}  \nonumber \\
\vec{\a}_{2n-1} &=& \vec{\varepsilon}_n - \vec{\delta}_n
{}~\quad \qquad \qquad \quad e_{2n-1}^+ = e_{n,3n+1} + e_{4n+1,2n} \nonumber \\
\vec{\a}_{2n} &=& \vec{\delta}_n \qquad \qquad ~~~~\qquad \qquad
e_{2n}^+ = e_{2n+1,4n+1} - e_{3n+1,2n+1}
\ena
We note that in Fig.$1$ the Dynkin diagram of the $B(n,n)$ superalgebra
can be obtained by folding the corresponding one of the $D(n+1,n)$ series.
The untwisted fermionic extension is obtained by adding to the
set of simple roots the lowest root
\EQ
\vec{\a}_0 = -\vec{\varepsilon}_1 - \vec{\delta}_1
\quad \qquad \qquad e_0^+ = e_{n+1,2n+2} +e_{3n+2,1}
\EN
As for the previous case, the generators for the negative roots are defined
as $e_j^- = e_{ba} \mp e_{dc}$, whenever the corresponding
positive root has generator $e_j^+ = e_{ab} \pm e_{cd}$.
An explicit realization for the $2n$--dimensional vectors
$\vec{\varepsilon}$~'s and $\vec{\delta}$~'s is
\bea
(\vec{\varepsilon}_j)_k &=& \delta_{kj} ~\quad \qquad \qquad \qquad j = 1,
\cdots ,n \nonumber \\
(\vec{\delta}_j)_k &=& i\delta_{k, j+n} \qquad \qquad \qquad j=1, \cdots ,n
\label{varep}
\ena
A basis for the Cartan subalgebra is
\bea
h_a &=& e_{a,a} - e_{a+n,a+n} ~\qquad \qquad \qquad \qquad \qquad a=1, \cdots
,n \nonumber \\
h_a &=& i(e_{a+n+1,a+n+1} - e_{a+2n+1,a+2n+1}) ~~~\qquad a = n+1, \cdots ,2n
\ena
Also in this case,
the Weyl vector is given by the expression in eq.(\ref{weyl}).

The calculation of the
weights of the fundamental representation gives the following result
\bea
\vec{\l}_j &=& -\vec{\l}_{4n+2-j}~~ =~~  \sum_{k=j}^{2n} \vec{\a}_k
\qquad ~~ j=1,\cdots , 2n \nonumber \\
\vec{\l}_{2n+1} &=& 0
\ena
where we have labelled the weights from the highest $\vec{\l}_1$ to the lowest
$\vec{\l}_{4n+1}$.

\vskip 10pt
\noindent
3) $A(n,n)$, $n \geq 1$

The fundamental representation of $A(n,n)$ is given in terms of
matrices with supertrace equal to zero in ${\cal M}(n+1|n+1;\cal{C})$.
This superalgebra is not simple since it contains
a one--dimensional invariant subalgebra generated by the element
${\bf I}_{2n+2}$. In the following we refer to $A(n,n)$ as the superalgebra
obtained by factoring out the invariant subalgebra.
This factorization is realized explicitly by identifying any two elements
of $A(n,n)$ whenever they differ by a multiple of the identity ${\bf
I}_{2n+2}$.
The rank of the algebra is $2n$ but, as a consequence of the fact that the
algebra is not simple, the number of simple roots exceeds the rank by one.
Only $2n$ of them
are independent, so that they can still be represented by vectors in a
$2n$--dimensional linear space. A realization of the fermionic roots
corresponding to the Dynkin diagram in Fig.1 can be given in terms of
vectors $\vec{\varepsilon}_i$, $\vec{\d}_j$ as in eq.(\ref{varep}).
We obtain for the roots and corresponding generators
\bea
\vec{\a}_1 &=& \vec{\varepsilon}_1 - \vec{\delta}_1
{}~\qquad \qquad \qquad \qquad \quad~~~ e_1^+ = \sqrt{2} e_{1,n+2}
\nonumber \\
\vec{\a}_2 &=&  \vec{\varepsilon}_1 + \vec{\delta}_1
{}~\qquad \qquad \qquad \qquad \quad~~~ e_2^+ = \sqrt{2} e_{n+2,2}
\nonumber \\
\vec{\a}_3 &=& -\vec{\varepsilon}_1 +\vec{\delta}_1 +\vec{\varepsilon}_2 -
\vec{\delta}_2 \qquad ~~~
{}~\qquad e_3^+ = \sqrt{2} e_{2,n+3} \nonumber \\
\vec{\a}_4 &=& \vec{\varepsilon}_2 + \vec{\delta}_2
{}~\qquad \qquad \qquad \qquad \quad~~~e_4^+ = \sqrt{2} e_{n+3,3} \nonumber \\
& \cdots & \qquad \qquad \qquad \qquad \qquad \qquad \quad~~~~
\cdots \nonumber \\
\vec{\a}_{2n-1} &=& -\vec{\varepsilon}_{n-1} +\vec{\delta}_{n-1} +
\vec{\varepsilon}_n - \vec{\delta}_n
\quad~~~
e_{2n-1}^+ = \sqrt{2} e_{n,2n+1} \nonumber \\
\vec{\a}_{2n} &=& \vec{\varepsilon}_n + \vec{\delta}_n
\qquad \qquad \qquad \qquad ~~\quad e_{2n}^+ = \sqrt{2} e_{2n+1,n+1}
\nonumber \\
\vec{\a}_{2n+1} &=& - \vec{\varepsilon}_n + \vec{\delta}_n
\quad \qquad \qquad ~~~\qquad e_{2n+1}^+ = \sqrt{2} e_{n+1,2n+2}
\ena
As noticed above the simple roots
are not all independent since $\sum_{j=0}^n \vec{\a}_{2j+1} = 0$.
The fermionic extension is realized by
\EQ
\vec{\a}_0 = -\sum_{j=1}^n (\vec{\varepsilon}_j + \vec{\delta}_j )
{}~~\qquad \qquad \qquad e_0^+ = \sqrt{2} e_{2n+2,1}
\EN
The generators corresponding to negative simple roots are
$e_{2j}^- =- (e_{2j}^+)^{\rm t}$,
$e_{2j+1}^- = (e_{2j+1}^+)^{\rm t}$, $j = 0, \cdots , n$.
A basis of the Cartan subalgebra is given by the following matrices
\bea
h_a &=& \sum_{j=1}^a e_{j,j} - e_{a+1,a+1} + \sum_{j=2}^a e_{n+j,n+j}
\qquad \qquad~~  a=1, \cdots ,n
\nonumber \\
h_a &=& i ( \sum_{j=1}^{a-n+1} e_{j,j} + \sum_{j=2}^{a-n} e_{n+j,n+j}
+ 2e_{a+1,a+1} )
\qquad a=n+1, \cdots ,2n
\ena
where we have identified
\EQ
e_{2n+2,2n+2} \equiv - \sum_{k=1}^{2n+1} e_{k,k}
\EN
as a consequence of the factorization of the invariant subalgebra.
The Weyl vector can be defined as half the sum of vectors dual
to the first $2n$ independent simple roots. We obtain
\EQ
\vec{\rho} = \frac{1}{4} \sum_{k=1}^n [(k+1) \vec{\varepsilon}_k
+ (k-1) \vec{\delta}_k]
\EN
Finally the weights of the fundamental representation, with the unique ordering
from the highest to the lowest one, are
\bea
\vec{\l}_{j} &=& \sum_{k=j}^{2n+1} \vec{\a}_k \qquad \qquad j=1,\cdots , 2n+1
\nonumber \\
\vec{\l}_{2n+2} &=&  0
\ena

\newpage


\begin{center}
\begin{tabular}{lclc}
\setlength{\unitlength}{1.5pt}
\begin{picture}(400,140)(-31,-90)
\thicklines

\put(-31,10){\makebox(0,0){$D(n+1,n):$}}
\put(0,10){\circle{8}}
\put(0,10){\makebox(0,0){$\times$}}
\put(-2,-1){\makebox(0,0){$\a_1$}}
\put(4,10){\line(1,0){15}}
\put(23,10){\circle{8}}
\put(23,10){\makebox(0,0){$\times$}}
\put(21,-1){\makebox(0,0){$\a_2$}}
\put(27,10){\line(1,0){7}}
\put(38,10){\line(1,0){7}}
\put(49,10){\line(1,0){7}}
\put(60,10){\circle{8}}
\put(60,10){\makebox(0,0){$\times$}}
\put(58,-1){\makebox(0,0){$\a_{2n-1}$}}
\put(64,10){\line(1,1){9}}
\put(64,10){\line(1,-1){9}}
\put(77,19){\circle{8}}
\put(77,19){\makebox(0,0){$\times$}}
\put(78,30){\makebox(0,0){$\a_{2n}$}}
\put(77,1){\circle{8}}
\put(77,1){\makebox(0,0){$\times$}}
\put(78,-10){\makebox(0,0){$\a_{2n+1}$}}
\put(76,5){\line(0,1){10}}
\put(78,5){\line(0,1){10}}

\put(-25,-54){\makebox(0,0){$B(n,n):$}}
\put(0,-54){\circle{8}}
\put(0,-54){\makebox(0,0){$\times$}}
\put(-2,-65){\makebox(0,0){$\a_1$}}
\put(4,-54){\line(1,0){15}}
\put(23,-54){\circle{8}}
\put(23,-54){\makebox(0,0){$\times$}}
\put(21,-65){\makebox(0,0){$\a_2$}}
\put(27,-54){\line(1,0){7}}
\put(38,-54){\line(1,0){7}}
\put(49,-54){\line(1,0){7}}
\put(60,-54){\circle{8}}
\put(60,-54){\makebox(0,0){$\times$}}
\put(58,-65){\makebox(0,0){$\a_{2n-1}$}}
\put(64,-52){\line(1,0){15}}
\put(64,-56){\line(1,0){15}}
\put(73,-54){\makebox(0,0){$>$}}
\put(83,-54){\circle*{8}}
\put(81,-65){\makebox(0,0){$\a_{2n}$}}

\put(-25,-120){\makebox(0,0){$A(n,n):$}}
\put(0,-120){\circle{8}}
\put(0,-120){\makebox(0,0){$\times$}}
\put(-2,-131){\makebox(0,0){$\a_1$}}
\put(4,-120){\line(1,0){15}}
\put(23,-120){\circle{8}}
\put(23,-120){\makebox(0,0){$\times$}}
\put(21,-131){\makebox(0,0){$\a_2$}}
\put(27,-120){\line(1,0){7}}
\put(38,-120){\line(1,0){7}}
\put(49,-120){\line(1,0){7}}
\put(60,-120){\circle{8}}
\put(60,-120){\makebox(0,0){$\times$}}
\put(58,-131){\makebox(0,0){$\a_{2n}$}}
\put(64,-120){\line(1,0){15}}
\put(83,-120){\circle{8}}
\put(83,-120){\makebox(0,0){$\times$}}
\put(81,-131){\makebox(0,0){$\a_{2n+1}$}}

\put(148,19){\line(1,-1){9}}
\put(148,1){\line(1,1){9}}
\put(144,19){\circle{8}}
\put(144,19){\makebox(0,0){$\times$}}
\put(136,16){\makebox{$1$}}
\put(144,1){\circle{8}}
\put(144,1){\makebox(0,0){$\times$}}
\put(136,-2){\makebox{$1$}}
\put(143,5){\line(0,1){10}}
\put(145,5){\line(0,1){10}}
\put(161,10){\circle{8}}
\put(161,10){\makebox(0,0){$\times$}}
\put(159,-1){\makebox{$2$}}
\put(165,10){\line(1,0){15}}
\put(184,10){\circle{8}}
\put(184,10){\makebox(0,0){$\times$}}
\put(182,-1){\makebox{$2$}}
\put(189,10){\line(1,0){7}}
\put(200,10){\line(1,0){7}}
\put(211,10){\line(1,0){7}}
\put(222,10){\circle{8}}
\put(222,10){\makebox(0,0){$\times$}}
\put(220,-1){\makebox{$2$}}
\put(226,10){\line(1,1){9}}
\put(226,10){\line(1,-1){9}}
\put(239,19){\circle{8}}
\put(239,19){\makebox(0,0){$\times$}}
\put(244,16){\makebox{$1$}}
\put(239,1){\circle{8}}
\put(239,1){\makebox(0,0){$\times$}}
\put(244,-2){\makebox{$1$}}
\put(238,5){\line(0,1){10}}
\put(240,5){\line(0,1){10}}

\put(148,-45){\line(1,-1){9}}
\put(148,-63){\line(1,1){9}}
\put(144,-45){\circle{8}}
\put(144,-45){\makebox(0,0){$\times$}}
\put(136,-48){\makebox{$1$}}
\put(144,-63){\circle{8}}
\put(144,-63){\makebox(0,0){$\times$}}
\put(136,-67){\makebox{$1$}}
\put(143,-59){\line(0,1){10}}
\put(145,-59){\line(0,1){10}}
\put(161,-54){\circle{8}}
\put(161,-54){\makebox(0,0){$\times$}}
\put(159,-65){\makebox{$2$}}
\put(165,-54){\line(1,0){15}}
\put(184,-54){\circle{8}}
\put(184,-54){\makebox(0,0){$\times$}}
\put(182,-65){\makebox{$2$}}
\put(189,-54){\line(1,0){7}}
\put(200,-54){\line(1,0){7}}
\put(211,-54){\line(1,0){7}}
\put(222,-54){\circle{8}}
\put(222,-54){\makebox(0,0){$\times$}}
\put(220,-65){\makebox{$2$}}
\put(226,-52){\line(1,0){15}}
\put(226,-56){\line(1,0){15}}
\put(233,-56){\makebox{$>$}}
\put(245,-54){\circle*{8}}
\put(243,-65){\makebox{$2$}}

\put(143,-120){\circle{8}}
\put(143,-120){\makebox(0,0){$\times$}}
\put(141,-131){\makebox{$1$}}
\put(147,-120){\line(1,0){15}}
\put(166,-120){\circle{8}}
\put(166,-120){\makebox(0,0){$\times$}}
\put(164,-131){\makebox{$1$}}
\put(170,-120){\line(1,0){7}}
\put(181,-120){\line(1,0){7}}
\put(192,-120){\line(1,0){7}}
\put(203,-120){\circle{8}}
\put(203,-120){\makebox(0,0){$\times$}}
\put(201,-131){\makebox{$1$}}
\put(207,-120){\line(1,0){15}}
\put(226,-120){\circle{8}}
\put(226,-120){\makebox(0,0){$\times$}}
\put(224,-131){\makebox{$1$}}
\put(147,-119){\line(2,1){33}}
\put(189,-102){\line(2,-1){33}}
\put(185,-102){\circle{8}}
\put(185,-102){\makebox(0,0){$\times$}}
\put(183,-97){\makebox{$1$}}

\end{picture}
\end{tabular}
\end{center}
\vspace{5cm}
\begin{center}
{\bf Figure 1:}
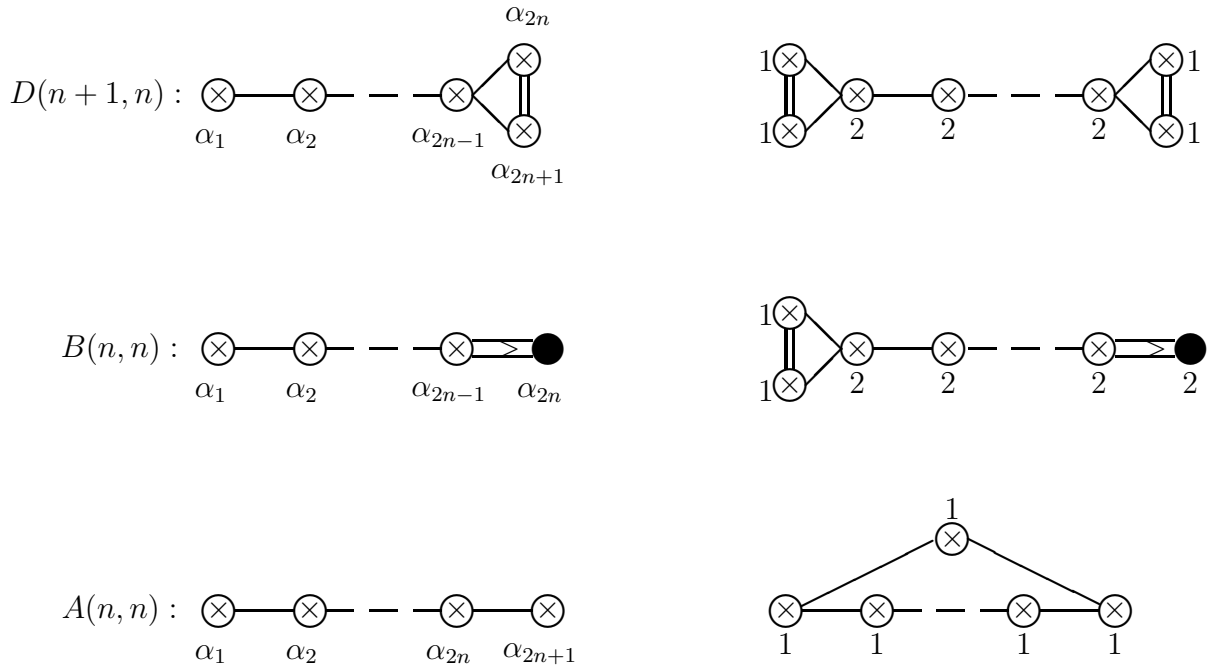
Fermionic Dynkin diagrams for Lie superalgebras and their affine
{}~~~~~~~~~~~~~~~~~~extensions. The crossed circles denote fermionic roots
with vanishing norm.~

The numbers in the affine Dynkin diagrams are the Ka$\check
{c}$ labels.~~~~~~~~~~~~~~~~~~~
\end{center}

\newpage


\begin{center}
\begin{tabular}{lclc}
\setlength{\unitlength}{1.5pt}
\begin{picture}(400,140) (-28,-95)
\thicklines

\put(13,10){\circle{26}}
\put(26,10){\line(1,0){20}}
\put(-20,10){\line(1,0){20}}
\put(13,-10){\makebox(0,0){a)}}
\put(-22,10){\makebox(0,0){$1$}}
\put(48,10){\makebox(0,0){$1$}}
\put(-1,16){\makebox(0,0){$4$}}
\put(-1,4){\makebox(0,0){$4$}}

\put(119,10){\circle{26}}
\put(132,10){\line(1,0){20}}
\put(86,10){\line(1,0){20}}
\put(119,-10){\makebox(0,0){b)}}
\put(84,10){\makebox(0,0){$1$}}
\put(154,10){\makebox(0,0){$1$}}
\put(105,16){\makebox(0,0){$1$}}
\put(105,4){\makebox(0,0){$2$}}

\put(225,10){\circle{26}}
\put(238,10){\line(1,0){20}}
\put(192,10){\line(1,0){20}}
\put(225,-10){\makebox(0,0){c)}}
\put(190,10){\makebox(0,0){$1$}}
\put(260,10){\makebox(0,0){$1$}}
\put(211,16){\makebox(0,0){$2$}}
\put(211,4){\makebox(0,0){$3$}}

\put(13,-50){\circle{26}}
\put(26,-50){\line(1,0){20}}
\put(-20,-50){\line(1,0){20}}
\put(13,-70){\makebox(0,0){d)}}
\put(-22,-50){\makebox(0,0){$2$}}
\put(48,-50){\makebox(0,0){$2$}}
\put(-1,-44){\makebox(0,0){$1$}}
\put(-1,-56){\makebox(0,0){$1$}}

\put(119,-50){\circle{26}}
\put(132,-50){\line(1,0){20}}
\put(86,-50){\line(1,0){20}}
\put(119,-70){\makebox(0,0){e)}}
\put(84,-50){\makebox(0,0){$2$}}
\put(154,-50){\makebox(0,0){$2$}}
\put(105,-44){\makebox(0,0){$3$}}
\put(105,-56){\makebox(0,0){$3$}}

\put(225,-50){\circle{26}}
\put(238,-50){\line(1,0){20}}
\put(192,-50){\line(1,0){20}}
\put(225,-70){\makebox(0,0){f)}}
\put(190,-50){\makebox(0,0){$2$}}
\put(260,-50){\makebox(0,0){$2$}}
\put(211,-44){\makebox(0,0){$4$}}
\put(211,-56){\makebox(0,0){$4$}}

\put(119,-110){\circle{26}}
\put(132,-110){\line(1,0){20}}
\put(86,-110){\line(1,0){20}}
\put(119,-130){\makebox(0,0){g)}}
\put(84,-110){\makebox(0,0){$2$}}
\put(154,-110){\makebox(0,0){$2$}}
\put(105,-104){\makebox(0,0){$1$}}
\put(105,-116){\makebox(0,0){$3$}}

\put(13,-170){\circle{26}}
\put(26,-170){\line(1,0){20}}
\put(-20,-170){\line(1,0){20}}
\put(13,-190){\makebox(0,0){h)}}
\put(-22,-170){\makebox(0,0){$3$}}
\put(48,-170){\makebox(0,0){$3$}}
\put(-1,-164){\makebox(0,0){$2$}}
\put(-1,-176){\makebox(0,0){$3$}}

\put(119,-170){\circle{26}}
\put(132,-170){\line(1,0){20}}
\put(86,-170){\line(1,0){20}}
\put(119,-190){\makebox(0,0){i)}}
\put(84,-170){\makebox(0,0){$3$}}
\put(154,-170){\makebox(0,0){$3$}}
\put(105,-164){\makebox(0,0){$4$}}
\put(105,-176){\makebox(0,0){$4$}}

\put(225,-170){\circle{26}}
\put(238,-170){\line(1,0){20}}
\put(192,-170){\line(1,0){20}}
\put(225,-190){\makebox(0,0){l)}}
\put(190,-170){\makebox(0,0){$3$}}
\put(260,-170){\makebox(0,0){$3$}}
\put(211,-164){\makebox(0,0){$1$}}
\put(211,-176){\makebox(0,0){$2$}}

\put(13,-230){\circle{26}}
\put(26,-230){\line(1,0){20}}
\put(-20,-230){\line(1,0){20}}
\put(13,-250){\makebox(0,0){m)}}
\put(-22,-230){\makebox(0,0){$4$}}
\put(48,-230){\makebox(0,0){$4$}}
\put(-1,-224){\makebox(0,0){$2$}}
\put(-1,-236){\makebox(0,0){$4$}}

\put(119,-230){\circle{26}}
\put(132,-230){\line(1,0){20}}
\put(86,-230){\line(1,0){20}}
\put(119,-250){\makebox(0,0){n)}}
\put(84,-230){\makebox(0,0){$4$}}
\put(154,-230){\makebox(0,0){$4$}}
\put(105,-224){\makebox(0,0){$1$}}
\put(105,-236){\makebox(0,0){$4$}}

\put(225,-230){\circle{26}}
\put(238,-230){\line(1,0){20}}
\put(192,-230){\line(1,0){20}}
\put(225,-250){\makebox(0,0){o)}}
\put(190,-230){\makebox(0,0){$4$}}
\put(260,-230){\makebox(0,0){$4$}}
\put(211,-224){\makebox(0,0){$3$}}
\put(211,-236){\makebox(0,0){$4$}}

\end{picture}
\end{tabular}
\end{center}
\vspace{10cm}

\begin{center}
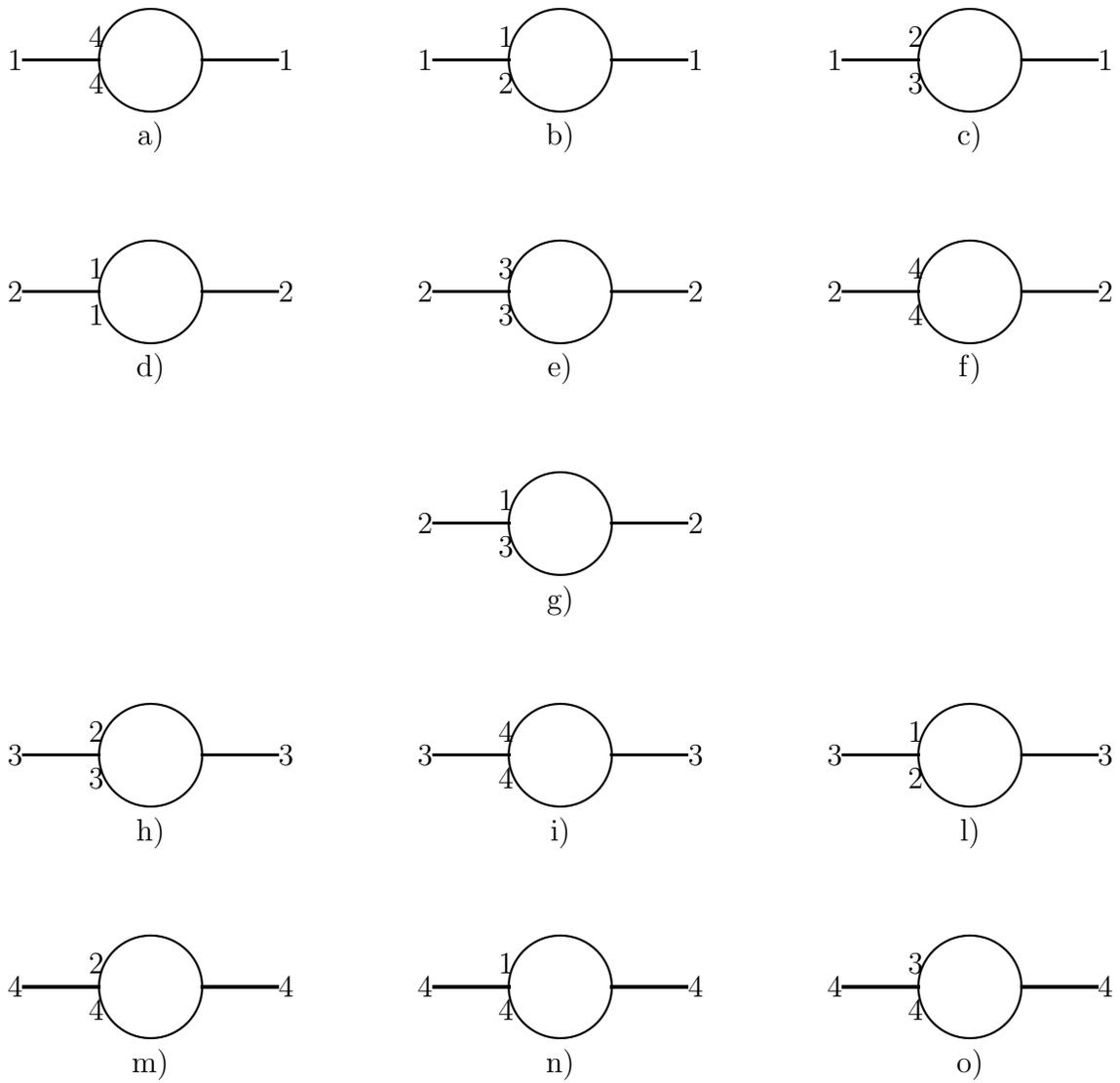
{\bf Figure 2:} Diagrams of one--loop mass corrections for
the $B^{(1)}(2,2)$ theory.

The numbers denote the various
particles.~~~~~~~~~~~~~~~~~~~~~~~~~~~~~~~~~~~~~~~
{}~~~~
\end{center}

\end{document}